\documentclass[12pt]{article}\usepackage[]{graphicx}\usepackage[]{xcolor}
\makeatletter
\def\maxwidth{ %
  \ifdim\Gin@nat@width>\linewidth
    \linewidth
  \else
    \Gin@nat@width
  \fi
}
\makeatother

\definecolor{fgcolor}{rgb}{0.345, 0.345, 0.345}

\usepackage{framed}
\makeatletter
 {\par\unskip\endMakeFramed%
 \at@end@of@kframe}
\makeatother

\definecolor{shadecolor}{rgb}{.97, .97, .97}
\definecolor{messagecolor}{rgb}{0, 0, 0}
\definecolor{warningcolor}{rgb}{1, 0, 1}
\definecolor{errorcolor}{rgb}{1, 0, 0}
\newenvironment{knitrout}{}{} 

\usepackage{alltt}
\usepackage{graphicx} 
\usepackage{amsmath}
\usepackage{amssymb} 
\usepackage{amsthm} 
\usepackage{dsfont} 
\usepackage[comma]{natbib} 
\usepackage{changes}
\usepackage{tikz}
\usetikzlibrary{positioning, shapes}
\allowdisplaybreaks
\newcommand{\bsc}{\boldsymbol{c}}
\newcommand{\bsx}{\boldsymbol{x}}
\newcommand{\bseta}{\boldsymbol{\eta}}
\newcommand{\xinfty}{x\rightarrow\infty}
\newcommand{\limx}{\lim_{\xinfty}}
\newcommand{\ninfty}{n\rightarrow\infty}
\newcommand{\limn}{\lim_{\ninfty}}
\newcommand{\indicator}[1]{\mathbb{I}_{\left\{#1\right\}}}
\newtheorem{theorem}{Theorem}

\newtheorem{lemma}[theorem]{Lemma}

\newcommand{\blind}{0}

\IfFileExists{upquote.sty}{\usepackage{upquote}}{}
\begin{document}

\def\spacingset#1{\renewcommand{\baselinestretch}%
{#1}\small\normalsize} \spacingset{1}

\if0\blind
{
  \title{\bf Degree distributions in networks: beyond the power law}
  \author{Clement Lee\thanks{Corresponding author. Email: clement.lee@newcastle.ac.uk}\\
    School of Mathematics, Statistics and Physics, Newcastle University\\
    and \\
    Emma F. Eastoe and Aiden Farrell\\
    Department of Mathematics and Statistics, Lancaster University}
  \maketitle
} \fi

\if1\blind
{
  \bigskip
  \bigskip
  \bigskip
  \begin{center}
    {\LARGE\bf Degree distributions in networks: beyond the power law}
\end{center}
  \medskip
} \fi

\bigskip
\begin{abstract}
The power law is useful in describing count phenomena such as network degrees and word frequencies. With a single parameter, it captures the main feature that the frequencies are linear on the log-log scale. Nevertheless, there have been criticisms of the power law, for example that a threshold needs to be pre-selected without its uncertainty quantified, that the power law is simply inadequate, and that subsequent hypothesis tests are required to determine whether the data could have come from the power law. We propose a modelling framework that combines two different generalisations of the power law, namely the generalised Pareto distribution and the Zipf-polylog distribution, to resolve these issues. The proposed mixture distributions are shown to fit the data well and quantify the threshold uncertainty in a natural way. A model selection step embedded in the Bayesian inference algorithm further answers the question whether the power law is adequate.

\end{abstract}

\noindent%
\vfill

\newpage
\spacingset{1.75} 

\section{Introduction} \label{sect.intro}
In statistics, the power law is a principle by which the relative change in the size of a random variable is proportional to the relative change in its frequency. The continuous distribution associated with the power law is the Pareto distribution, of which the probability density function (PDF) is
\begin{align}
  f(z)=\frac{(\alpha-1)}{u}\left(\frac{z}{u}\right)^{-\alpha},\qquad z>u, \label{eqn.intro_power_freq}
\end{align}
where $u>0$ is the threshold, and $\alpha>1$ is the exponent parameter. The analogous distribution for discrete data is the Zipf distribution, with probability mass function (PMF)
\begin{align}
  p(x) =\frac{x^{-\alpha}}{\zeta(\alpha,u+1)},\qquad x=u+1,u+2,\ldots, \label{eqn.intro_zipf_pmf}
\end{align}
where $u$ is a non-negative integer, and $\zeta(\alpha,z)=\displaystyle\sum_{i=0}^{\infty}(z+i)^{-\alpha}$ is the Hurwitz zeta function. Whilst the form of this distribution with $u=0$ has many other names including the zeta distribution and the discrete Pareto distribution \citep{jkk05}, we use the name Zipf distribution for consistency with  \cite{vpl22}. The Zipf distribution has been applied to data in various fields including quantitative linguistics \citep{bgsl12}, casualty numbers \citep{bgdsj09,friedman15,gillespie17}, and city sizes \citep{newman05,csn09}. 

In this paper, we focus on a specific class of discrete data which is routinely associated with the power law -- the (in-)degrees of a network. Examples include networks of links on the World-Wide Web \citep{ba99a,ajb99,fff99}, social networks \citep{lo14,varga15}, coauthorship and citation networks \citep{price76,tw14,athr20}, and retweet counts \citep{bsz15, mmnb17}. Networks for which the degree distribution follows a power law are described as scale-free.

When studying the degree distribution of a network, the behaviour of the large (in-)degrees is of most interest because this can identify the generative mechanism of the network. Specifically, the popular preferential attachment (PA) generative model \citep{ba99a}, which extends the cumulative advantage process \citep{price76}, generates networks for which the right tail of the degree distribution follows the power law. Therefore, determining whether the empirical right tail of the degree distribution of a network follows a power law can inform the choice of model for the network generating mechanism.

Despite the popularity of the PA model, many have argued that application of the power law to degree distributions is inappropriate, that scale-free networks are limiting and that the PA model is inadequate. For example, \cite{swm05} found that sampling from scale-free networks does not result in (smaller) scale-free networks. \cite{sp12} questioned the ubiquity of the power law in different disciplines, whilst \cite{bc19} found that scale-free networks are rare in reality, and for most networks the log-normal distribution is sufficient. While \cite{vvvk19} and \cite{asvw20} criticised the latter, the findings have been supported by, for example, \cite{tw14}, \cite{so18}, and \cite{athr20}. 

This paper aims to address these, and other, concerns about the use of the power law to model degree distributions. To motivate our work, we examine some real-world examples of network degree distributions using two popular graphical diagnostics. From Equation \ref{eqn.intro_zipf_pmf} the logarithm of the Zipf distribution, $\log{}(p(x))$, is linear with $\log{}(x)$. Hence the power law is a suitable model for a degree distribution if the empirical frequency, or equivalently the empirical survival function, has a linear relationship with the data  on the log-log scale.  Figure \ref{fig:plot-example-data} displays both diagnostics for four examples, two from the KONECT database \citep{kunegis13} and two from \cite{vpl22}. First observe that, for all data sets, a large proportion of the nodes in the network have very small degrees. For example, for both data sets on the left, the values $x=1,\ldots,10$ make up 81\% of the data. In other words, if the right tail of the degree distribution were defined to be all values above a high threshold, then it would consist of almost all the unique values in the data set.

\begin{knitrout}
\definecolor{shadecolor}{rgb}{0.969, 0.969, 0.969}\color{fgcolor}\begin{figure}[!h]

{\centering \includegraphics[width=\maxwidth]{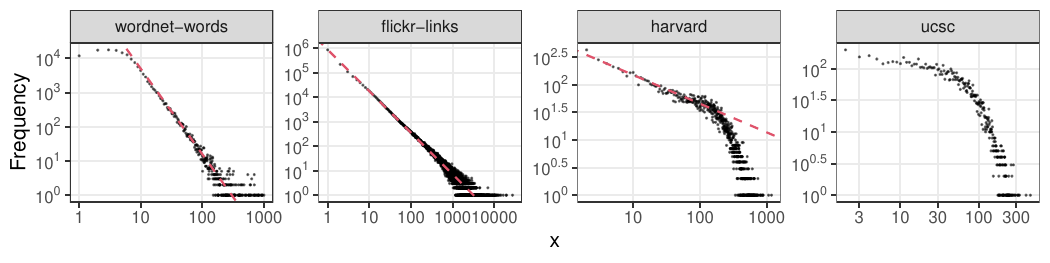} 
\includegraphics[width=\maxwidth]{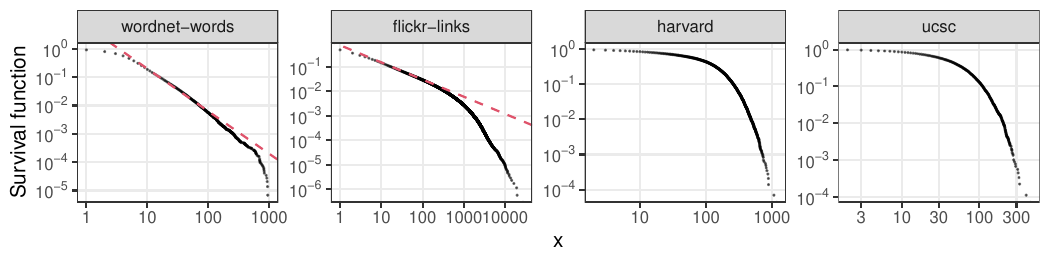} 

}

\caption[Empirical frequencies (top) and survival function (bottom) on the log-log scale of four data sets]{Empirical frequencies (top) and survival function (bottom) on the log-log scale of four data sets. Barring the rightmost data set, a red dashed line with slope $-\alpha$ in the top plot implies a dashed line with slope $-(\alpha-1)$ in the bottom plot, and suggests a power law with exponent $\alpha$.}\label{fig:plot-example-data}
\end{figure}

\end{knitrout}

Now consider the linearity of the plots. The ``wordnet-words'' data shows piecewise linearity, with the left and right tails showing different behaviour. In this case, it might be appropriate to fit the Zipf distribution to the largest degrees, however this requires a subjective choice of a cut-off point, or threshold, with which to identify the largest degrees which, in turn, will impact inference. Further, by construction, the models are fitted using different subsets of the data meaning standard model selection tests cannot be used to identify the threshold. \cite{csn09} and \cite{gillespie15} suggested using the Kolmogorov-Smirnov statistic, but this requires additional procedures beyond fitting the Zipf distribution whilst not resolving  the loss of information.

The ``flickr-links'' data is a case in which the Zipf distribution might fit the body of the data well but would be insufficiently flexible for the right tail. This data also clearly demonstrates the utility of the survival plot over the frequency plot. That the right tail should be separately modelled aligns with the work of \cite{vvvk19}, who suggested the use of regularly varying distributions in such cases. Since such distributions have heavy right tails, this approach struggles with data with light tails and, further, still requires a subset of the data to be identified prior to performing inference.

The body and tail of the ``harvard'' data also differ. In this case, when the right tail is removed, a truncated Zipf distribution with $\alpha\leq 1$ is appropriate. Aside from the lack of fit to the right tail, such a power law is not appropriate for the whole support of the data since the resulting distribution does not have a proper PMF. 
Lastly, as shown by the ``ucsc'' data, the power law may simply not be appropriate. In this case, the survival function is neither fully, nor piecewise, linear.

In summary, the Zipf distribution is too simplistic to satisfactorily capture the empirical degree distribution of many real-world data sets, and even when the power law is appropriate for a subset of the data, there is neither a systematic, objective way of choosing this subset nor a way to quantify the uncertainty of this choice. We address both issues by proposing a modelling and inference framework that uses a mixture of discrete  distributions to accommodate survival curves with a piecewise nature. While  \cite{jp21} proposed a similar approach, their model is a mixture of Zipf distributions with the additional restriction that the same exponent parameter $\alpha$ be used for all components, resulting in very limited flexibility. In contrast, we use a mixture of Zipf-type and extreme value distributions, which have disjoint supports and model the body and the tail(s) separately. More importantly, analogous to existing continuous counterparts in the literature, our approach allows estimation of the threshold(s) between the component distributions. Using a Bayesian inference approach permits both estimation of threshold uncertainty and model selection. The latter embedded naturally in the inference algorithm, thus removing the need for further hypothesis testing.

The rest of the paper is as follows. In Section \ref{sect.dist}, distributions related to the power law and extreme value theory are introduced, along with the characterisation of the asymptotic tail behaviour of the former. Two of these distributions form the mixture model, which is proposed in Section \ref{sect.model}. The inference algorithm is outlined in Section \ref{sect.inf}. Applications to real-world data are presented in Section \ref{sect.app}. The discussion in Section \ref{sect.discussion} concludes the paper.

\section{Distributions and properties} \label{sect.dist}

\subsection{The truncated Zipf-polylog distribution} \label{sect.zipf}
The \textit{truncated} Zipf-polylog$(\alpha,\theta,u,w)$ (TZP) distribution has the PMF
\begin{align}
  p_{\text{TZP}}(x;\alpha,\theta,u,w) =\frac{x^{-\alpha}\theta^x}{\displaystyle\sum_{k={w+1}}^u k^{-\alpha}\theta^k},\qquad x=w+1,w+2,\ldots,u, \label{eqn.tzp_pmf}
\end{align}
where $\theta\in(0,1]$, $u$ and $w$ are integers such that $u>w\geq0$, and $\alpha\in\mathds{R}$ when $\theta\in(0,1)$ or $u<\infty$, and $\alpha>1$ when $\theta=1$ and $u=\infty$. Several special cases of this distribution are summarised in Figure \ref{fig:dist-relationship}, including the Zipf-polylog (ZP) distribution \citep{vpl22}. In what follows, the \textit{polylog} distribution refers to the ZP distribution when $\theta\in(0,1)$, and consequently the ZP$(\alpha,\theta,w)$ distribution is the disjoint union of the Zipf$(\alpha,w)$ and polylog$(\alpha,\theta,w)$ distributions. Similarly, the TZP$(\alpha,\theta,u,w)$ distribution is the disjoint union of the Zipf-Mandelbrot$(\alpha,u,w)$ (ZM) and \textit{truncated polylog}$(\alpha,\theta,u,w)$ (TP) distributions. 

For completeness, the Yule-Simon distribution \citep{yule25}, which also implies an approximate power law in the right tail, is included in Figure \ref{fig:dist-relationship}. Note that the Yule-Simon$(\rho)$ distribution arises when $X|\theta\sim\text{Geometric}(\theta)$ and $\theta\sim\text{Beta}(\rho,1)$; \cite{ja22} proposed a generative network model based on this relationship, however they did not see the connection to the Yule-Simon distribution.

\begin{figure}[!h]
\centering
\begin{tikzpicture}[
roundnode/.style={ellipse, draw=green!60, fill=green!5, very thick, minimum size=7mm, font={\footnotesize}},
squarednode/.style={rectangle, draw=red!60, fill=red!5, very thick, minimum size=5mm, font={\footnotesize}},
align=center
]
\matrix[row sep=0.5cm, column sep=2.5cm, nodes=draw]
{
  \node[draw=none, fill=none] {}; &
  \node[squarednode] (ys)     {Yule-Simon$(\rho)$}; &
  \node[draw=none, fill=none] {}; \\
  \node[squarednode] (m)      {Zipf-Mandelbrot\\$(\alpha,u,w)$ /\\power law on\\$\{w+1,\ldots,u\}$}; &
  \node[squarednode] (zipf)   {Zipf$(\alpha,w)$ / zeta /\\power law on\\$\{w+1,\ldots\}$}; &
  \node[roundnode]   (p)      {Pareto$(\alpha,u)$ /\\power law on\\$(u,\infty)$}; \\
  \node[squarednode] (trun)   {\textbf{Truncated}\\ \textbf{Zipf-polylog}\\$(\alpha,\theta,u,w)$}; &
  \node[squarednode] (poly)   {Zipf-polylog\\$(\alpha,\theta,w)$}; &
  \node[roundnode]   (gpd)    {Generalised Pareto\\$(u,\sigma_0+\xi(u-\mu),\xi)$}; \\
  \node[squarednode] (geom)   {Geometric$(\theta)$}; &
  \node[squarednode] (log)    {Logarithmic$(\theta)$ /\\log-series}; &
  \node[squarednode] (igpd)   {\textbf{Integer}\\\textbf{generalised Pareto}\\$(u,\sigma_0+\xi(u-\mu),\xi)$}; \\
};

\draw[->] (gpd.north) -- (p.south) node [left, midway, font={\footnotesize}] {$\sigma_0=\xi\mu,\,\xi=1/(\alpha-1)$};
\draw[<->, dotted, thick] (p.west) -- (zipf.east) node [above, midway, font={\footnotesize}] {conceptually\\similar};
\draw[<->, dotted, thick] (ys.south) -- (zipf.north) node [left, midway, font={\footnotesize}] {conceptually similar};
\draw[->, dotted, thick] (igpd.north) -- (gpd.south) node [left, midway, font={\footnotesize}] {discretises};
\draw[->] (m.east) -- (zipf.west) node [above, midway, font={\footnotesize}] {$u=\infty$};
\draw[->] (poly.north) -- (zipf.south) node [left, midway, font={\footnotesize}] {$\theta=1$};
\draw[->] (poly.south) -- (log.north) node [left, midway, font={\footnotesize}] {$\alpha=1,$\\$w=0$};
\draw[->] (poly.south west) -- (geom.north east) node [left, midway, font={\footnotesize}] {$\alpha=0,w=0$};
\draw[->] (trun.east) -- (poly.west) node [above, midway, font={\footnotesize}] {$u=\infty$};
\draw[->] (trun.north) -- (m.south) node [left, midway, font={\footnotesize}] {$\theta=1$};
\draw[->, dotted, thick, shorten >=1mm]
  (geom.west) -- ++(-3mm,0mm)
  |- node [near end, yshift=-0.75em, font={\footnotesize}] {$\theta\sim\text{Beta}(\rho,1)$} (ys);
\end{tikzpicture}
\caption{Relationships between various continuous (green oval) and discrete (red rectangular) distributions. A solid arrow from $A$ to $B$ means $B$ is a special case of $A$. The meanings of various dashed arrows are given in the figure. Distributions that imply the power law are in the second row, while those that form the proposed mixture distributions are bolded.} \label{fig:dist-relationship}
\end{figure}

\subsection{From the generalised Pareto distribution to its discrete version} \label{sect.evt}
We next look at how the generalised Pareto (GP) distribution generalises the Pareto distribution, how it is used as a statistical model, and its discretisation. We begin with its PDF
\begin{align}
  g(z)=\left\{\begin{array}{ll}
  \displaystyle\frac{1}{\sigma_0}\left[1+\xi\frac{z-\mu}{\sigma_0}\right]_+^{-1/\xi-1}, & \xi\neq0,\\
  \displaystyle\frac{1}{\sigma_0}\exp\left[-\frac{z-\mu}{\sigma_0}\right], & \xi=0,
  \end{array}\right.\nonumber
\end{align}
where $A_+:=\max\{A,0\}$, $\mu$ is the location parameter, $\sigma_0>0$ is the scale paramater, and $\xi$ is the tail index or shape parameter. The support is $x\geq\mu$ when $\xi\geq0$, and $\mu\leq{}x\leq\mu-\sigma_0/\xi$ when $\xi<0$. There is no loss in generality in working primarily with the upper expression as the lower expression is its limit as $\xi\rightarrow0$. If a random variable $Z$ follows the GP$(\mu,\sigma_0,\xi)$ distribution exactly, the distribution of $Z|Z>u$ has the PDF
\begin{align}
  g_u(z)=\displaystyle\frac{1}{\sigma_0+\xi(u-\mu)}\left[1+\frac{\xi(z-u)}{\sigma_0+\xi(u-\mu)}\right]_+^{-1/\xi-1},\label{eqn.intro_gpd_z_u}
\end{align}
which is the GP$(u,\sigma_0+\xi(u-\mu),\xi)$ distribution. To see how the GP distribution generalises the Pareto distribution, note that when $\xi>0$ and $\sigma_0=\xi\mu$, $g_u(z)$ is equivalent to the density in Equation \ref{eqn.intro_power_freq} with $\alpha=1/\xi+1$.  In practice, for identifiability, the GP distribution in Equation \ref{eqn.intro_gpd_z_u} is re-parametrised as $(\phi_u,\sigma,\xi)$ where $\phi_u:=\Pr(Z>u)=\left[1+\xi\frac{u-\mu}{\sigma_0}\right]_{+}^{-1/\xi}$ is the exceedance probability, and $\sigma=\sigma_0-\xi\mu$. For this and other practical considerations when fitting the GP distribution to real-world data, see, for example, \cite{coles01}.

The practical importance of the GP distribution comes from the work of \cite{pickands75}, who showed that for $Z\sim{}F$ where $F$ is an arbitrary continuous CDF and high $u$, the GP distribution approximates the conditional tail $\Pr(Z>z|Z>u)$. Further, the GP distribution is the only distribution with this capability. The approximation follows from asymptotic results which show that for almost all absolutely continuous distributions $F$, the conditional distribution in Equation \ref{eqn.intro_gpd_z_u} holds exactly in the limit as $u\rightarrow\infty$. These results support the use of the GP distribution in data analysis as a model for observations above a high threshold. We will look at how this threshold is chosen in Section \ref{sect.model}.

As the GP distribution is a continuous distribution, it is not appropriate to apply it directly to discrete data. Whilst \cite{vvvk19} proposed adding uniform noise to  integer-valued data before estimating $\xi$ using the continuous distribution, a more elegant solution is to use the integer-GP (IGP) distribution \citep{pgs14, reft18}, the discretised version of the GP distribution. Specifically, if $Z$ follows the GP distribution with PDF $g(z)$, then $X=\lceil{}Z\rceil$ follows the IGP distribution with the same parameters and has PMF $p(x)=\displaystyle\int_{x-1}^{x}g(z)dz$. The case where $X=\lfloor{}Z\rfloor$ has been considered by \cite{hds17}. The relationships between the Pareto, GP and IGP distributions are summarised in the rightmost column of Figure \ref{fig:dist-relationship}. Note that the IGP distribution has no direct relationship with the TZP distribution, even though both are discrete extensions of power law distributions.

\subsection{Domain of attraction of Zipf-polylog} \label{sect.doa}
The behaviour of the right tail of a continuous probability distribution $F_0$ can be characterised by the maximum domain of attraction (DoA) of the distribution. Specifically, $F_0$ is in the DoA of an extreme value distribution $H$ if there exists $a_n>0$, $b_n\in\mathds{R}$ such that $\limn\left|F_0^n(a_nx+b_n)-H(x)\right|=0,$ where $H$ must be a negative Weibull, Gumbel, or Fr\'{e}chet distribution. There is a one-to-one connection between this result and the GP distribution tail approximation: when the right tail of $F_0$ is approximated by the GP distribution with $\xi<0$, $\xi=0$, and $\xi>0$, then $F_0$ is in the DoA of the negative Weibull, Gumbel, and Fr\'{e}chet distributions, respectively. 

While the unified DoA of the three extreme value distributions covers a wide range of common continuous distributions, many common discrete distributions, including the geometric, do not belong to a DoA according to the definition above. To address this, \cite{shimura12} introduced the concept of recovery to a DoA. If a discrete distribution $F$ is the discretisation of a continuous distribution $F_0$ and $F_0$ is in a DoA, then $F$ is said to be recoverable to the same DoA. For example, the geometric distribution is a discretisation of the exponential distribution, which is in the Gumbel DoA, therefore the geometric distribution is recoverable to the Gumbel DoA. 

As the ZP distribution is the disjoint union of the polylog and Zipf distributions, understanding the right tail hebaviour of the ZP distribution requires us to first characterise the tail behaviour of both these distributions. Specifically, the DoA to which each of the polylog and Zipf distributions is recoverable can be used to determine the behaviour of the equivalent IGP tail approximation.

For the polylog distribution, the key quantity for  recovery is $$\Omega(F,x):=\left(\log\frac{S(x+1)}{S(x+2)}\right)^{-1}-\left(\log\frac{S(x)}{S(x+1)}\right)^{-1},$$ where $S(x)=1-F(x)$ is the survival function of the discrete distribution concerned. \cite{shimura12} showed that if $\displaystyle\limx\Omega(F,x)=0$, then $F$ is (uniquely) recoverable to the Gumbel DoA. This turns out to be the case for the polylog$(\alpha,\theta,w)$ distribution; see Appendix \ref{appendix.polylog} for the proof. Consequently, the IGP distribution that approximates the polylog$(\alpha,\theta,w)$ distribution has $\xi=0$ always.

The Zipf distribution requires the result that if $\displaystyle\limx\Omega(F,x)=\xi>0$, then the discrete distribution is in the DoA of the Fr\'{e}chet distribution with positive tail index $\xi$ \citep{shimura12}. For the Zipf$(\alpha,w)$ distribution, this limit is $1/(\alpha-1)$; see Appendix \ref{appendix.zipf} for the proof. Consequently, the IGP distribution that approximates the right tail of the Zipf$(\alpha,w)$ distribution  has $\xi=1/(\alpha-1)$. This mirrors the result for the continuous counterpart (Pareto distribution) established in, for example, \cite{vvvk19}.

Given these two results, it follows that the IGP approximation to the right tail of the ZP$(\alpha,\theta,w)$ distribution has $\xi=0$ when $\theta\in(0,1)$, and $\xi=1/(\alpha-1)$ when $\theta=1$. This means the ZP distribution is unlikely to fit both the body and the right tail of a discrete data set well, as its implied tail index, denoted by $\xi_{\text{ZP}}$, can take only one of the two values. Using the IGP distribution for the right tail better captures the tail behaviour.

\section{Mixture model} \label{sect.model}

\subsection{Threshold selection}
To motivate our model, we begin with an approach to threshold selection when applying the GP distribution to continuous data. As mentioned in Section \ref{sect.evt}, the GP distribution is appropriate for observations above a pre-selected high threshold $u$, regardless of the underlying generating distribution. An example of threshold selection for network and/or power-law-related data is \cite{vvvk19}. A more principled approach \citep{coles01} is to first fit the GP distribution over different values of $u$, assess the plots of the maximum likelihood estimates (MLEs) of $\sigma$ and $\xi$, and choose the threshold above which the parameter estimates look stable, after considering their uncertainty. Whilst useful, this still requires a subjective decision, and the uncertainty in $u$ is not accounted for. \cite{mtv23} proposed an objective procedure to select a single threshold that minimises the difference between the empirical and fitted Q-Q plots. However, the uncertainty around the single threshold is not considered in these approaches.

A solution to threshold selection where its uncertainty is considered is to use an extreme value mixture distributions; see \cite{sm12} for a comprehensive review of continuous mixture distributions, with numerical routines for fitting these distributions available in the R package evmix \citep{hs18}. An extreme value mixture distribution combines a GP distribution for the observations above $u$, and another distribution $H$ for those below $u$, while allowing $u$ to vary as a parameter and therefore to be estimated. Specifically, assuming $Y$ is a positive continuous random variable with PDF $h$ and CDF $H$, we consider another positive random variable $Z$ which equals $Y$ if $Z\leq u$, where $u$ is a positive threshold. Conditional on $Z>u$, $Z$ follows the GP$(\phi_u,\sigma,\xi)$ distribution. The PDF of this mixture distribution for $Z$ is
\begin{align}
  f(z)=\left\{\begin{array}{ll}
  (1-\phi_u)\times\displaystyle\frac{h(z)}{H(u)}, & z\leq u,\\
  \phi_u\times g_u(z), & z > u,
  \end{array}\right.
\end{align}
where $h$ and $g_u$ are referred to as the bulk and tail distributions, respectively. Both $\phi_u$ and $H(u)$ are required for $f$ to be a proper density, i.e., to integrate to 1.

\subsection{Discrete extreme value mixture distributions} \label{sect.mix2}
To adapt the extreme value mixture model for integer-valued data, the discrete counterpart to the GP distribution, the IGP distribution, should be used for the tail. Specification of the bulk distribution is less stringent. In the case of degree distributions, the power law, or a generalisation thereof, is usually both appropriate and desirable for the bulk of a degree distribution, and hence the natural candidate is the TZP distribution. 

A random variable $X$ follows the TZP-IGP$(\alpha,\theta_{\text{mix}},u,\sigma,\xi)$ distribution if its PMF is
\begin{align}
  p_2(x) =\left\{\begin{array}{cl}
  (1-\phi_u) p_{\text{TZP}}(x;\alpha_{\text{mix}},\theta_{\text{mix}},u,w), & x=w+1,w+2,\ldots,u, \\
  \phi_u\left[G_u(x;\sigma,\xi)-G_u(x-1;\sigma,\xi)\right], & x=u+1,u+2,\ldots
  \end{array}\right.\label{eqn.mix2_pmf}
\end{align}
where $p_{\text{TZP}}(x)$ is the PMF of the TZP distribution in Equation \ref{eqn.tzp_pmf}, and
\begin{align}
  G_u(z;\sigma,\xi)=1-\displaystyle\left[1+\frac{\xi(z-u)}{\sigma+\xi u}\right]_+^{-1/\xi}\nonumber
\end{align}
is the CDF of the GP distribution, derived from integrating $g_u(z)$ in Equation \ref{eqn.intro_gpd_z_u} under the current parameterisation. The ``$2$'' in $p_2(x)$ stands for 2 components in this mixture distribution, which is a weighted sum of the TZP and IGP distributions with weights $1-\phi_u$ and $\phi_u$, respectively.

\begin{knitrout}
\definecolor{shadecolor}{rgb}{0.969, 0.969, 0.969}\color{fgcolor}\begin{figure}[!h]

{\centering \includegraphics[width=0.49\linewidth]{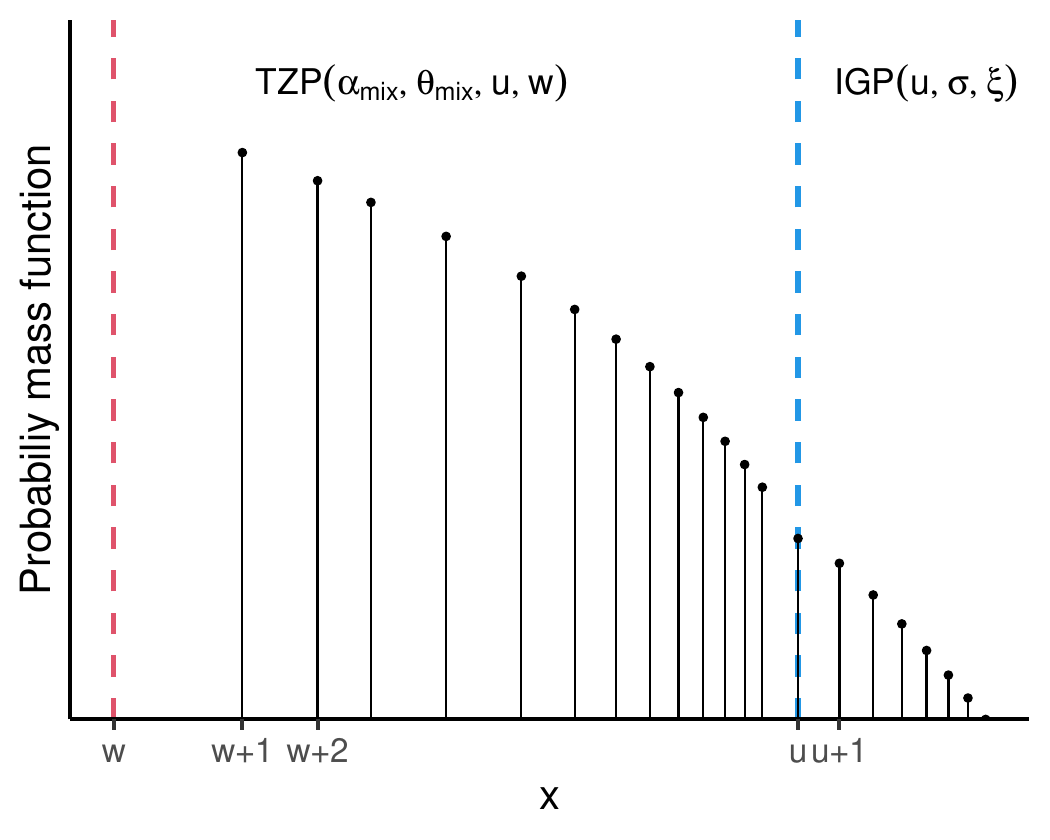} 
\includegraphics[width=0.49\linewidth]{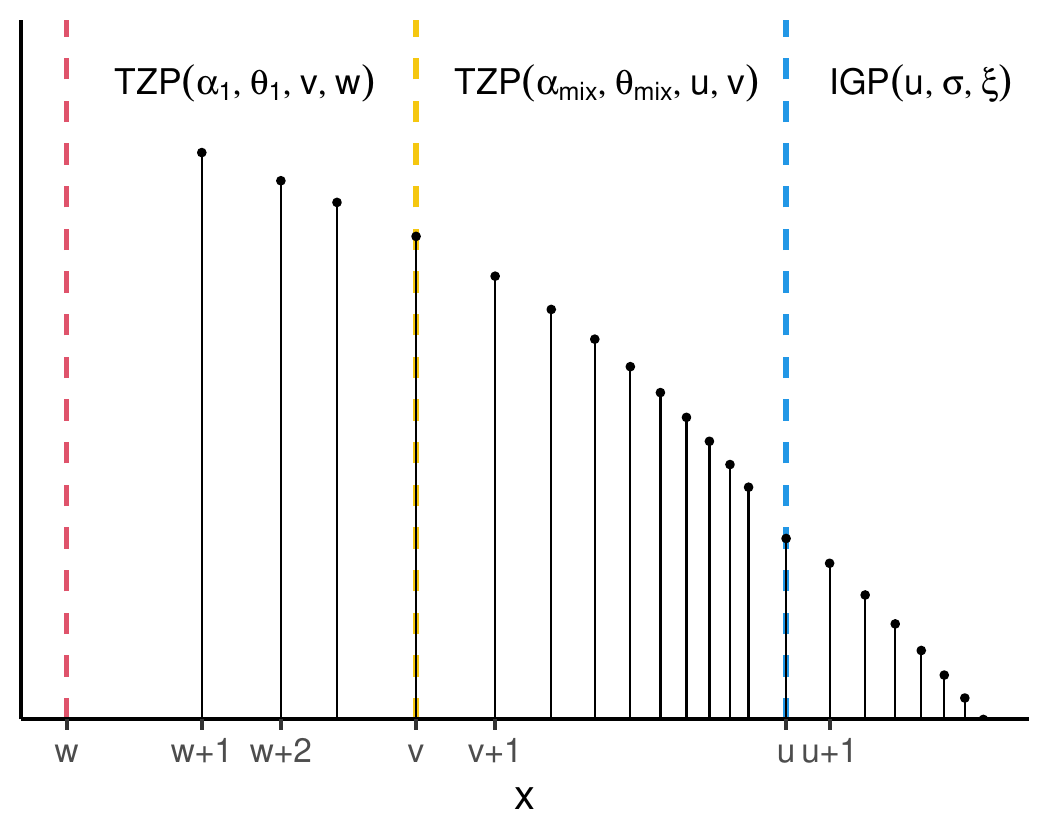} 

}

\caption[Schematic representations of the 2-component (left) and 3-component (right) mixture distributions]{Schematic representations of the 2-component (left) and 3-component (right) mixture distributions. Spacing between consecutive integers is not drawn to scale.}\label{fig:plot-schematic}
\end{figure}

\end{knitrout}

One point regarding the parametrisation is that $u$ is used here instead of $\phi_u$, as in Section \ref{sect.evt}. There is no issue with this, since there is a one-to-one relationship between the two parameters, with $\phi_u$  the empirical proprotion of exceedances once $u$ is given; see Section \ref{sect.like}. Consequently, either $\phi_u$ or $u$, but not both, can be a free parameter.

The number of components in the finite mixture modelling framework is not limited to 2; indeed a TZP-IGP mixture might still be inadequate. We also present the 3-component counterpart, where two different TZP distributions are used for $\{w+1,w+2,\ldots,v\}$ and $\{v+1,v+2,\ldots,u\}$, with $v$ an additional lower threshold. Mathematically, a random variable $X$ follows the TZP-TZP-IGP$(\alpha_1,\theta_1,v,\alpha_{\text{mix}},\theta_{\text{mix}},u,\sigma,\xi)$ distribution if its PMF is
\begin{align}
  \begin{split}
  p_3(x) =&~\phi_1\indicator{x\in\{w+1,\ldots,v\}}p_{\text{TZP}}(x;\alpha_1,\theta_1,v,w) \\
         &+ (1-\phi_1-\phi_u)\indicator{x\in\{v+1,\ldots,u\}}p_{\text{TZP}}(x;\alpha_{\text{mix}},\theta_{\text{mix}},u,v) \\
         &+ \phi_u\indicator{x\in\{u+1,\ldots\}}[G_u(x;\sigma,\xi)-G_u(x-1;\sigma,\xi)],\qquad x=w+1,w+2,\ldots
  \end{split}
\end{align}
where $\indicator{A}$ is the indicator function, taking value $1$ if event $A$ occurs and $0$ otherwise. We  call the 2-component and 3-component mixtures the TZP-IGP and TZP-TZP-IGP distributions, respectively; their schematic representations are in Figure \ref{fig:plot-schematic}. The former can be viewed as a disjoint union of the TP-IGP (when $\theta_{\text{mix}}\in(0,1)$) and ZM-IGP (when $\theta_{\text{mix}}=1$) distributions. Similarly, the latter is a disjoint union of the TZP-TP-IGP and TZP-ZM-IGP distributions.

Finally, we define the \textit{implied} tail index $\xi_{\text{mix}}=\indicator{\theta_{\text{mix}}=1}/(\alpha_{\text{mix}}-1)$, as discussed in Section \ref{sect.doa}, \textit{had the bulk TZP distribution been extended to encompass the right tail} and provided that $\alpha_{\text{mix}}>1$. Its difference with $\xi$ and $\xi_{\text{ZP}}$, defined in Section \ref{sect.doa}, will be key when showing the adequacy of the mixture distributions in Section \ref{sect.app}.

\subsection{Likelihood} \label{sect.like}
It is common for degree distributions to have multiple observations with the same integer value. For a sample of size $n$, and $m$ unique values $(x_1,x_2,\ldots,x_m)$ with counts $(c_1,c_2,\ldots,c_m)$ such that $\displaystyle\sum_{i=1}^mc_i=n$, the likelihood for the 2-component mixture distribution of Equation \ref{eqn.mix2_pmf} is
\begin{align}
  \begin{split}
  L_2(\bseta_2|\bsx,\bsc)=\prod_{i=1}^mp_2(x_i)^{c_i} &=(1-\phi_u)^{n-n_u}\phi_u^{n_u}\times\prod_{i:w<x_i\leq{}u}p_{\text{TZP}}(x_i;\alpha_{\text{mix}},\theta_{\text{mix}},u,w)^{c_i}\\
  &\quad\times\prod_{i:x_i>u}\left[G_u(x_i;\sigma,\xi)-G_u(x_i-1;\sigma,\xi)\right]^{c_i},
  \end{split}
    \label{eqn.mix2_like}
\end{align}
where $\bseta_2:=(w,\alpha_{\text{mix}},\theta_{\text{mix}},u,\sigma,\xi)$ is the parameter vector, and $n_u=\displaystyle\sum_{i=1}^mc_i\indicator{x_i>u}$ is the number of exceedances of $u$. The parameters $(\alpha_{\text{mix}}, \theta_{\text{mix}})$ and $(\sigma,\xi)$ describe the bulk and tail distributions, respectively. By definition of the model, for a given data set and for any given value of $u$, $\phi_u$ is completely determined. The converse is also true. We propose taking $u$ to be the free parameter, in which case $\phi_u$ can be estimated by its MLE $n_u/n$, i.e. the empirical proportion of exceedances, due to the factorisation of likelihood \eqref{eqn.mix2_like}. This echoes the point on the relationship between $u$ and $\phi_u$ in Section \ref{sect.mix2}, and is standard practice in extreme value methods \citep[for example]{coles01}. The likelihood for the 3-component mixture follows directly:
\begin{align*}
  \begin{split}
  L_3(\bseta_3|\bsx,\bsc)&=\phi_1^{n_1}\phi_2^{n-n_1-n_u}\phi_u^{n_u}\times\prod_{i:w<x_i\leq{}v}p_{\text{TZP}}(x_i;\alpha_1,\theta_1,v,w)^{c_i} \\
  &\quad\times\prod_{i:v<x_i\leq{}u}p_{\text{TZP}}(x_i;\alpha_{\text{mix}},\theta_{\text{mix}},u,v)^{c_i}\prod_{i:x_i>u}\left[G_u(x_i;\sigma,\xi)-G_u(x_i-1;\sigma,\xi)\right]^{c_i},
  \end{split}
\end{align*}
where $\bseta_3:=(w,\alpha_1,\theta_1,v,\alpha_{\text{mix}},\theta_{\text{mix}},u,\sigma,\xi)$. Similar to the 2-component mixture, when evaluating $L_3(\bseta_3|\bsx,\bsc)$, $\phi_1$ and $\phi_u$ are replaced by $n_1/n$ and $n_u/n$, respectively, where $n_1=\displaystyle\sum_{i=1}^mc_i\indicator{w<x_i\leq{}v}$.

Note that the likelihoods above assume that the degrees arise independently, which might not be realistic for network generative models such as the PA model. We shall stick to this assumption, however, as modelling the dependence between the degrees is out of the scope of this paper.

\section{Inference} \label{sect.inf}
We now outline the Bayesian inference framework, with emphasis on the model selection in order to decide whether the body of the degree distribution follows the power law. The 2-component mixture is used to illustrate the main steps of carrying out inference; similar steps apply to the 3-component mixture.

We choose $w$ to be pre-determined instead of inferred, as we would like it to be as low as possible to utilise as much data as possible in the inference. As, according to Equation \ref{eqn.mix2_pmf}, the support is $w+1,w+2,\ldots$, setting $w=0$ is a natural choice as the model will be suitable for all positive integers in the data. However, we take $w=1$ for consistency across data sets which do not have $1$s. Furthermore, for data sets that do have $1$s, including them usually worsens the fit. Consequently, only the $1$s must be discarded before fitting the mixture distribution.

\subsection{Profile likelihood of $u$} \label{sect.profile}
The frequentist framework provides a starting point of statistical inference. For fixed $u$, the MLEs of $\alpha_{\text{mix}}$ and $\theta_{\text{mix}}$ can be obtained by maximising $\displaystyle\prod_{i:x_i\leq{}u}p_{\text{TZP}}(x_i;\alpha_{\text{mix}},\theta_{\text{mix}},u,w)^{c_i}.$ Similarly, the MLEs of $\sigma$ and $\xi$ can be obtained by maximising $\displaystyle\prod_{i:x_i>u}\left[G_u(x_i;\sigma,\xi)-G_u(x_i-1;\sigma,\xi)\right]^{c_i}.$ Substituting the MLEs into Equation \ref{eqn.mix2_like} gives the profile log-likelihood, and hence the MLE, of $u$. While this approach is computationally efficient, the uncertainty around $u$ cannot be quantified by using the asymptotic behaviour of the MLE, as the required regularity conditions do not hold -- the profile log-likelihood usually has multiple discontinuities  and is therefore not differentiable. Furthermore, the profile log-likelihood tends to be quite flat, making it difficult to separate the tail from the bulk. 

In contrast, the Bayesian approach allows us to obtain the full joint posterior of $\bseta_2$ and quantify the uncertainty around $u$ via its marginal posterior. However, as $u$ is discrete, sampling from its posterior using all the values in the support is inefficient. Instead, we propose using its profile log-likelihood to identify the set of ``most probable'' values prior to applying the Bayesian inference algorithm. See Appendix \ref{appendix.profile} for further details. Note that the Bayesian approach does not remove the aforementioned possibility of a flat log-likelihood (or log-posterior density). Rather, it allows a better visualisation and quantification of the threshold uncertainty, which can vary across data sets; see the results in Section \ref{sect.app}, in particular Figure \ref{fig:plot-fitted-surv}.

\subsection{Priors} \label{sect.prior}
We assign independent and relatively uninformative priors as follows: $\alpha_{\text{mix}}\sim{}\text{N}(0,10^2)$, $\sigma\sim{}\text{Gamma}(1,0.01)$, and $\xi\sim{}\text{N}(0,10^2)$,
where $\text{N}(m,s^2)$ denotes the Gaussian distribution with expectation $m$ and variance $s^2$, and $\text{Gamma}(a,b)$ the Gamma distribution with shape parameter $a$ and rate parameter $b$. The prior for $\theta_{\text{mix}}$ is delayed to Section \ref{sect.model_select} to facilitate the model selection procedure.

The support of $u$ varies between data sets and it seems sensible to specify the prior for $u$ indirectly via a prior for $\phi_u$. However, as previously noted, the nature of degree distributions is such that a few small values  take up a large proportion of the data.  Further, a prior on  $(0,1)$ is equally undesirable as $\phi_u$ must be sufficiently small for the IGP distribution to approximate the tail well. Consequently, we consider the proportion of \textit{unique} exceedances of $u$, denoted by $\psi_u$. Similar to $\phi_u$, once $u$ is known, its empirical value can be computed as $\displaystyle\sum_{i=1}^m\indicator{x_i>{}u}/m$. We assume $\psi_u\sim\text{U}(0.001,0.9)$ \textit{a priori} and independent of the other priors, where $\text{U}(a,b)$ denotes the continuous uniform distribution with support $[a,b]$.

\subsection{Markov chain Monte Carlo and model selection} \label{sect.model_select}
According to Bayes theorem, $\pi(\bseta_2|\bsx,\bsc)=\displaystyle\frac{\pi(\bseta_2,\bsx,\bsc)}{\pi(\bsx,\bsc)}\propto\pi(\bseta_2,\bsx,\bsc)=L_2(\bseta_2|\bsx,\bsc)~\pi(\bseta_2)$, where $\pi(\bsx,\bsc)$ is the marginal likelihood and $\pi(\bseta_2)$ is the joint prior for $\bseta_2$. Essentially, the posterior for $\bseta_2$ is, up to a proportionality constant, the product of the likelihood and the prior for $\bseta_2$. As the proportionality constant is computationally intractable, the joint posterior can be estimated using a Markov chain Monte Carlo (MCMC) algorithm; see Appendix \ref{appendix.mcmc} for the details of the algorithm used in the application in Section \ref{sect.app}.

As discussed in Section \ref{sect.intro}, in the context of degree distributions, it is of interest to ascertain whether the distribution follows the power law, even if only in the body of the distribution. Under the mixture distribution, a power law in the body is implied when $\theta_{\text{mix}}=1$, and therefore a simple test would be to look at the (empirical) proportion of $1$'s in the marginal posterior distribution for $\theta_{\text{mix}}$. However, as $\theta_{\text{mix}}$ is  continuous, there is a zero probability of drawing $\theta_{\text{mix}}$ exactly equal to $1$ in the MCMC samples. Nor is it sufficient to consider $\theta_{\text{mix}}$ in a small interval close to $1$ since it was shown in Section \ref{sect.doa} that $\theta_{\text{mix}}\in(0,1)$ and $\theta_{\text{mix}}=1$  imply very different tail behaviours.

We resolve this issue by viewing the test between $\theta_{\text{mix}}\in(0,1)$ and $\theta_{\text{mix}}=1$ as a model selection between the TP-IGP and ZM-IGP distributions. The resulting model selection procedure requires neither that the models are nested nor that asymptotic results hold, and is embedded naturally in the MCMC algorithm \citep{cc95}. 

To implement the algorithm, we first complete the prior specification by assigning to $\theta_{\text{mix}}$ a spike-and-slab prior: $\theta_{\text{mix}}\sim{}(1-\gamma)\,\text{Beta}(1,1)+\gamma\,\delta_1(\cdot)$, where $\gamma\in(0,1)$ is pre-specified, $\text{Beta}(a,b)$ is the beta distribution with its common parametrisation, and $\delta_x(\cdot)$ is a discrete measure concentrated at $x$. This spike-and-slab prior, which is also assumed to be independent of the priors for other parameters, is common in model/variable selection \citep{ir05}. However, instead of working with this prior directly, it is useful to consider the equivalent formulation, where we introduce the \textit{model} variable $M:=\indicator{\theta_{\text{mix}}=1}$, assign prior probabilities $\Pr(M=1):=\gamma$ and $\Pr(M=0):=1-\gamma$, and specify $\theta_{\text{mix}}\sim\delta_1(\cdot)$ and $\theta_{\text{mix}}\sim\text{Beta}(1,1)$ when $M=1$ and $M=0$, respectively. Finally, we sample $M$ and $\bseta_2$ jointly in the MCMC algorithm, details of which are in Appendix \ref{appendix.mcmc}. The proportions of $1$s and $0$s in the output are the estimates of the posterior probabilities of the two models, denoted by $\hat{\Pr}(M=1|\bsx,\bsc)$ and $\hat{\Pr}(M=0|\bsx,\bsc)$, respectively. We then compute the Bayes factor $B_{10}=\displaystyle\frac{\hat{\Pr}(M=1|\bsx,\bsc)}{\hat{\Pr}(M=0|\bsx,\bsc)}\left/\frac{\Pr(M=1)}{\Pr(M=0)}\right.$; this is the evidence of ``the body of the data follows the power law'' against ``the body of the data does not follow the power law''.

For the 3-component mixture,  model selection takes place between TZP-TP-IGP ($\theta_{\text{mix}}\in(0,1)$) and TZP-ZM-IGP ($\theta_{\text{mix}}=1$), as whether the body between $v$ and $u$ follows the power law is of more interest than whether the left tail below $v$ follows the power law.

\section{Application} \label{sect.app}

In this section, we show the goodness-of-fit of the proposed distributions to a number of data sets and assess the suitability of the power law in each case. Of the data sets, most are undirected, unweighted networks from the KONECT database, omitting those with clear, and potentially artificial, jumps in the degree distributions. Also included are four data sets available in the \verb`R` package poweRlaw \citep{gillespie15}, including two sets of casualty numbers in armed conflicts (``us-americans'', ``native-americans'') \citep{bgdsj09,friedman15,gillespie17}, and two sets of word frequencies (``swiss-prot'', ``moby-dick'') investigated by \cite{bgsl12} and \cite{newman05}, respectively. Whilst not degree distributions, whether they follow the power law has been investigated by these references. Lastly, we also consider the directed dependencies of \verb`R` packages (``cran-dependencies''), available in the package crandep \citep{lee23}. As mentioned in Section \ref{sect.inf}, where a data set contains $1$s these have been removed.

For each data set, we first fit the 2-component mixture with the 3-component mixture fitted only if the former appears visually inadequate. We do not formally test between the mixtures as this is very computationally expensive. For comparison, we also fit the ZP distribution across the whole domain of the degree distribution, simultaneously carrying out model selection between the polylog $(\theta_{\text{ZP}}\in(0,1))$ and Zipf $(\theta_{\text{ZP}}=1)$ distributions. Again, the comparison between the best mixture and the selected ZP distribution is conducted visually, since formal testing would add considerable computational complexity.

\begin{knitrout}
\definecolor{shadecolor}{rgb}{0.969, 0.969, 0.969}\color{fgcolor}\begin{figure}

{\centering \includegraphics[width=0.24\linewidth]{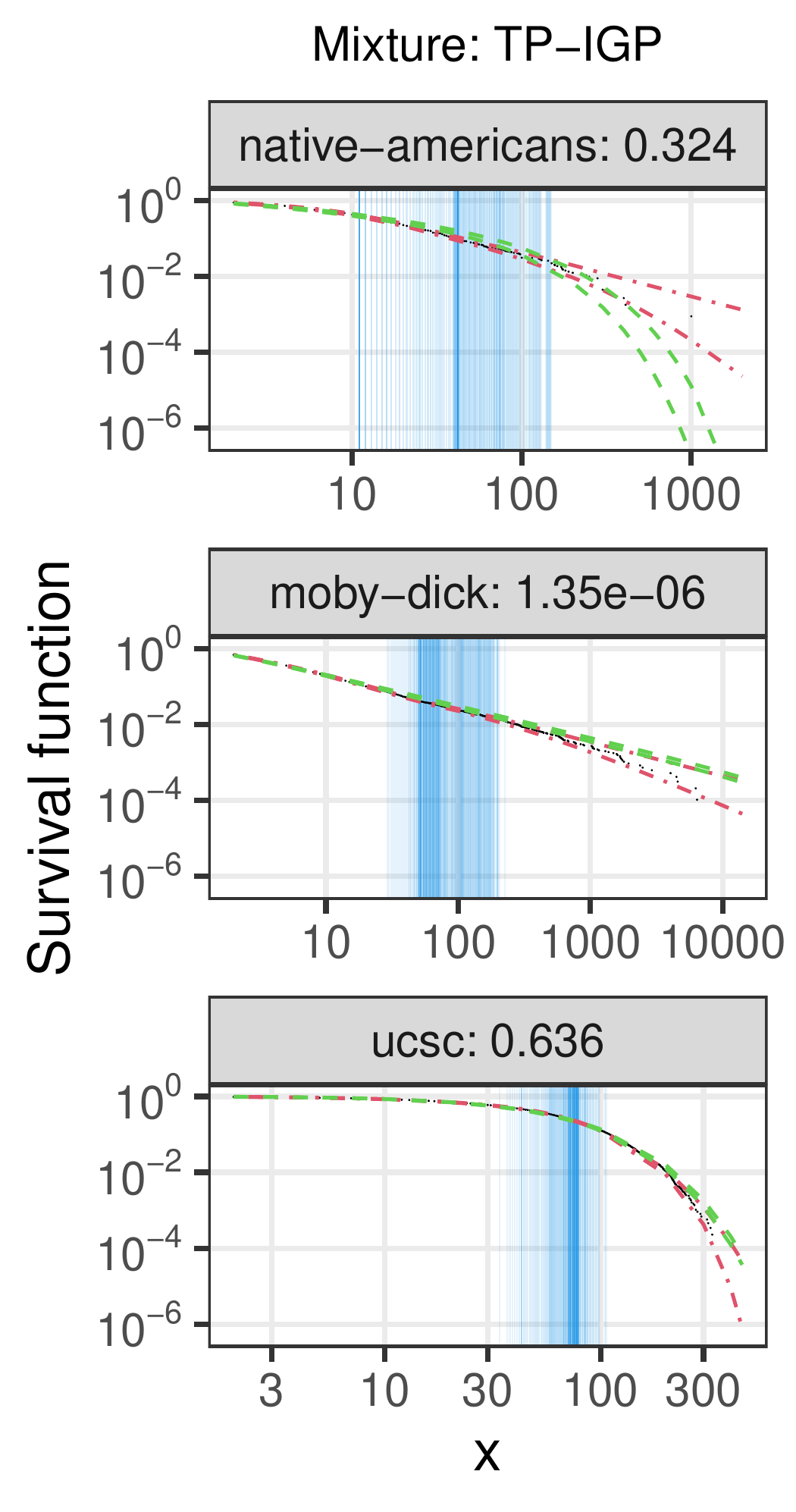} 
\includegraphics[width=0.24\linewidth]{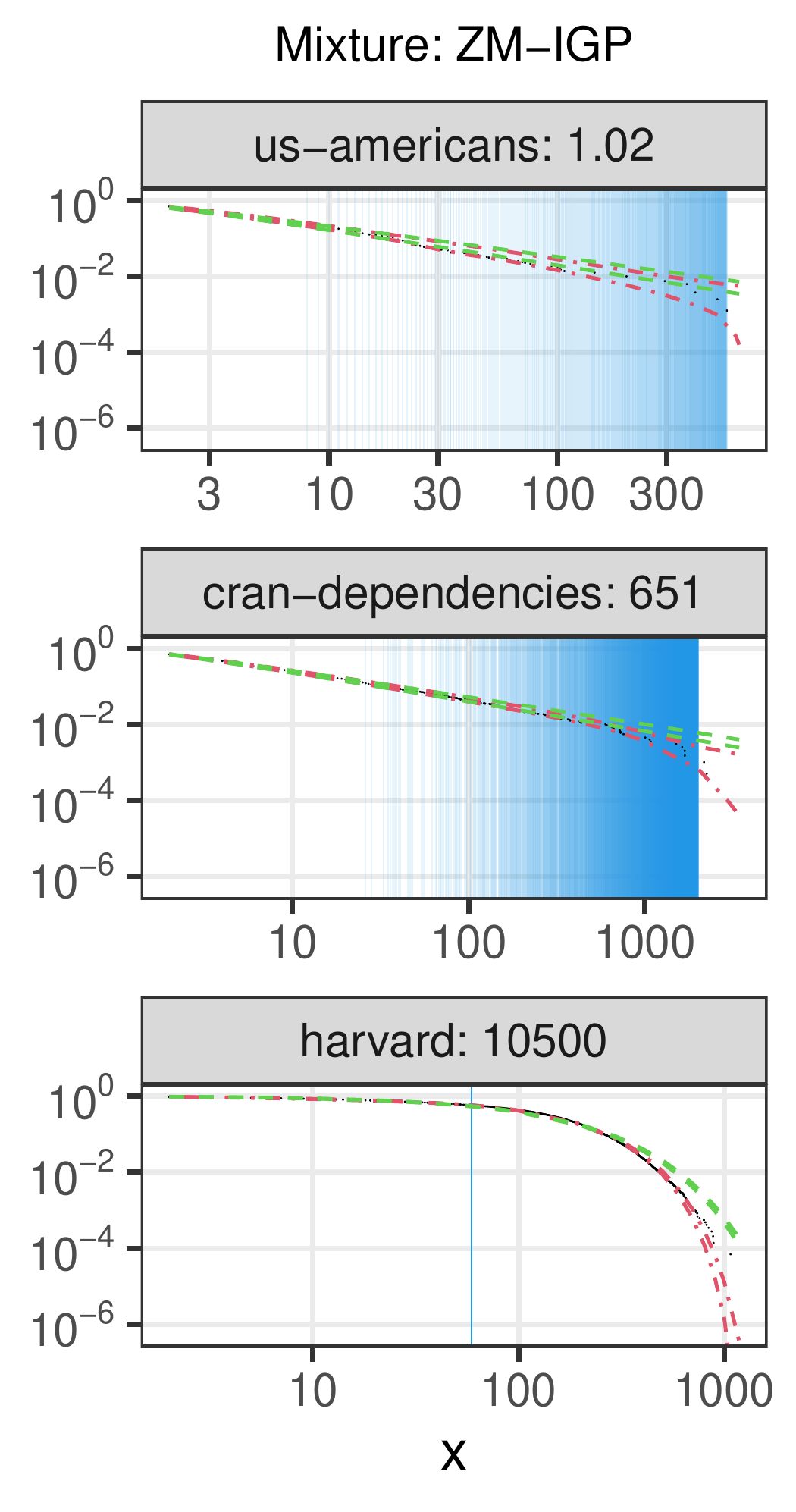} 
\includegraphics[width=0.24\linewidth]{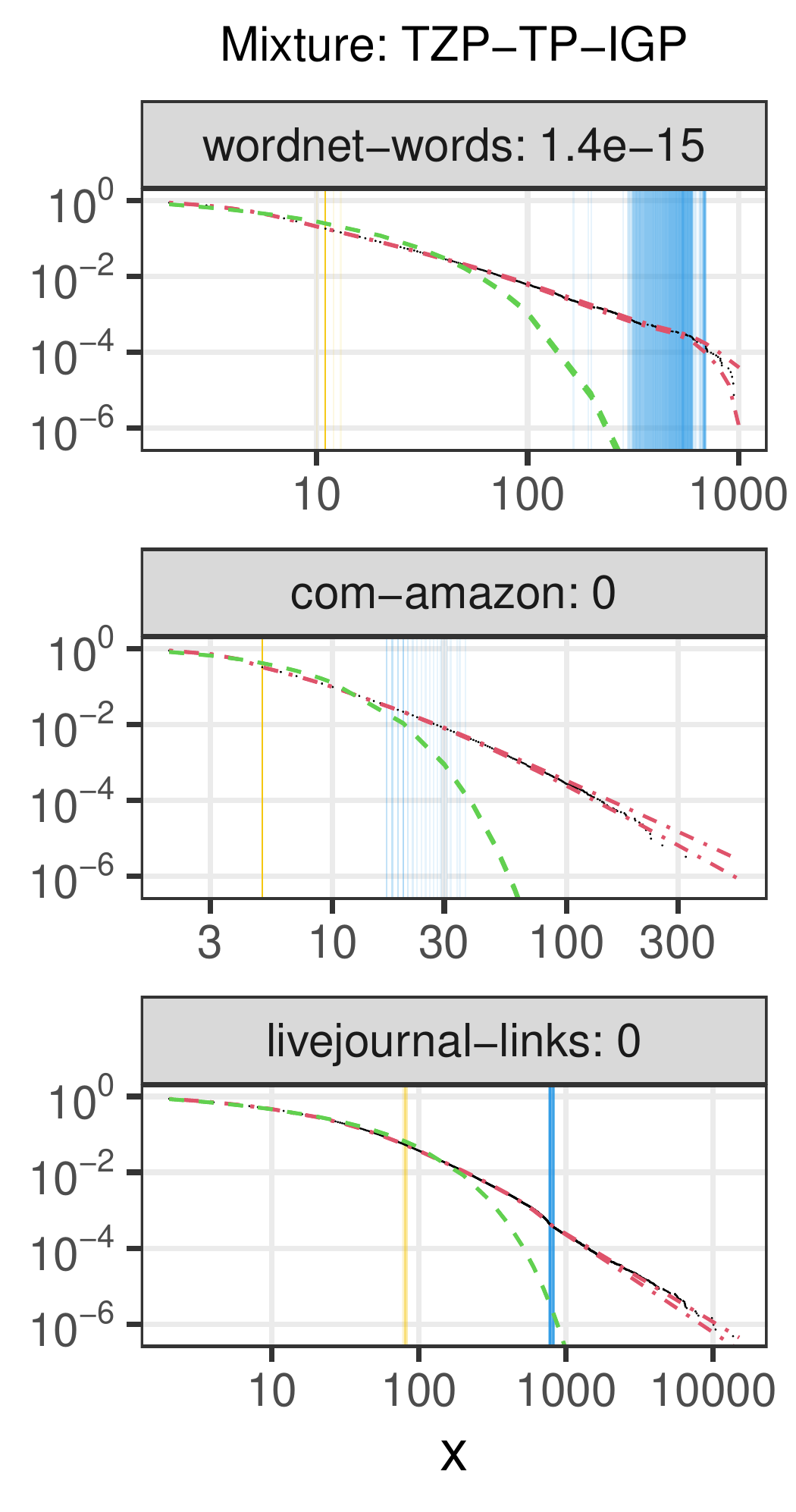} 
\includegraphics[width=0.24\linewidth]{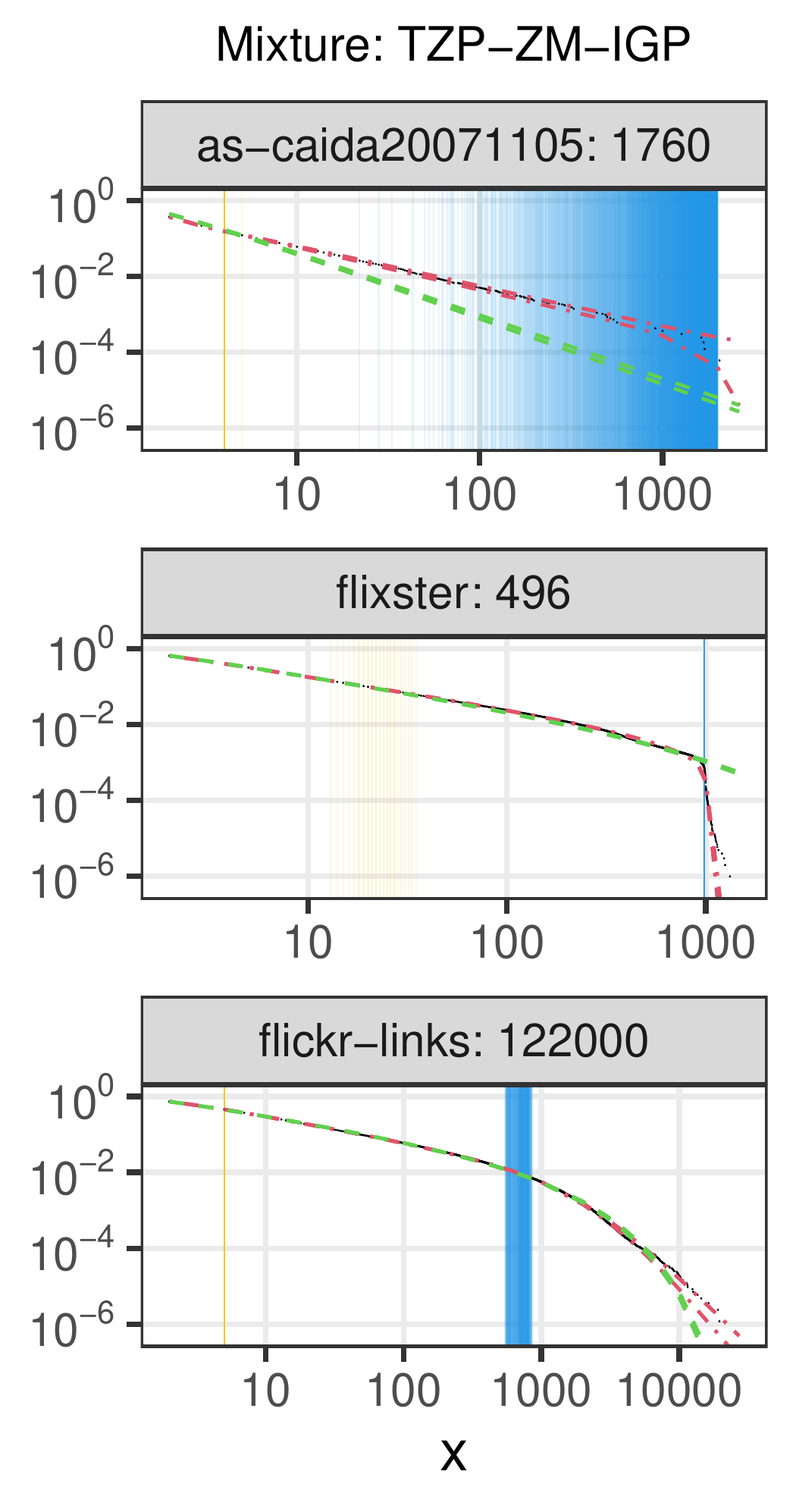} 

}

\caption[Survival function with credible intervals for the selected mixture (red) and ZP (green) distributions]{Survival function with credible intervals for the selected mixture (red) and ZP (green) distributions. The blue and yellow bands represent the posterior of $u$ and $v$, respectively, for the mixture distribution fits. The number beside the data set name is the Bayes factor (to 3 s.f.) for $\theta_{\text{mix}}=1$ relative to $\theta_{\text{mix}}\in(0,1)$.}\label{fig:plot-fitted-surv}
\end{figure}

\end{knitrout}

\begin{knitrout}
\definecolor{shadecolor}{rgb}{0.969, 0.969, 0.969}\color{fgcolor}\begin{figure}

{\centering \includegraphics[width=0.24\linewidth]{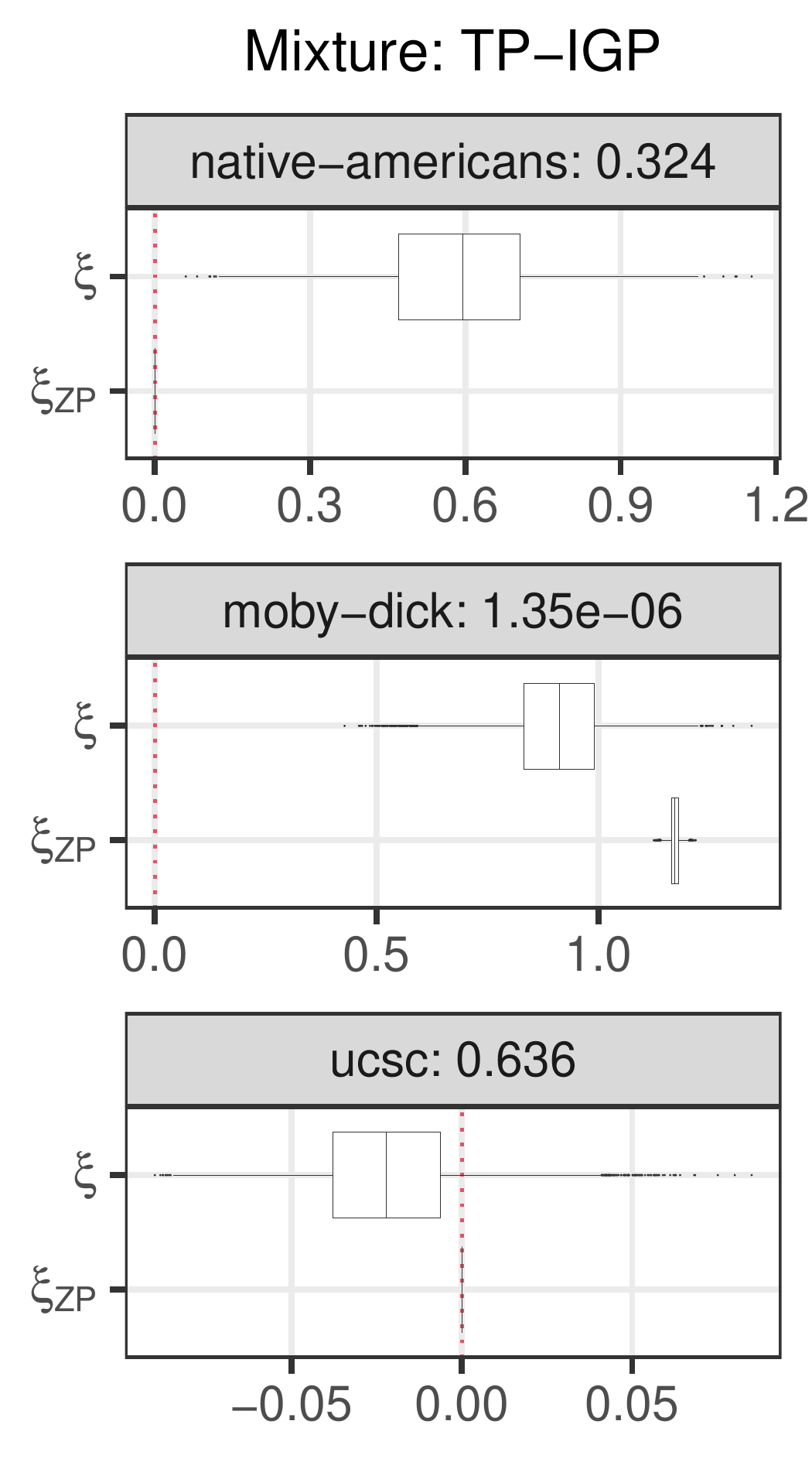} 
\includegraphics[width=0.24\linewidth]{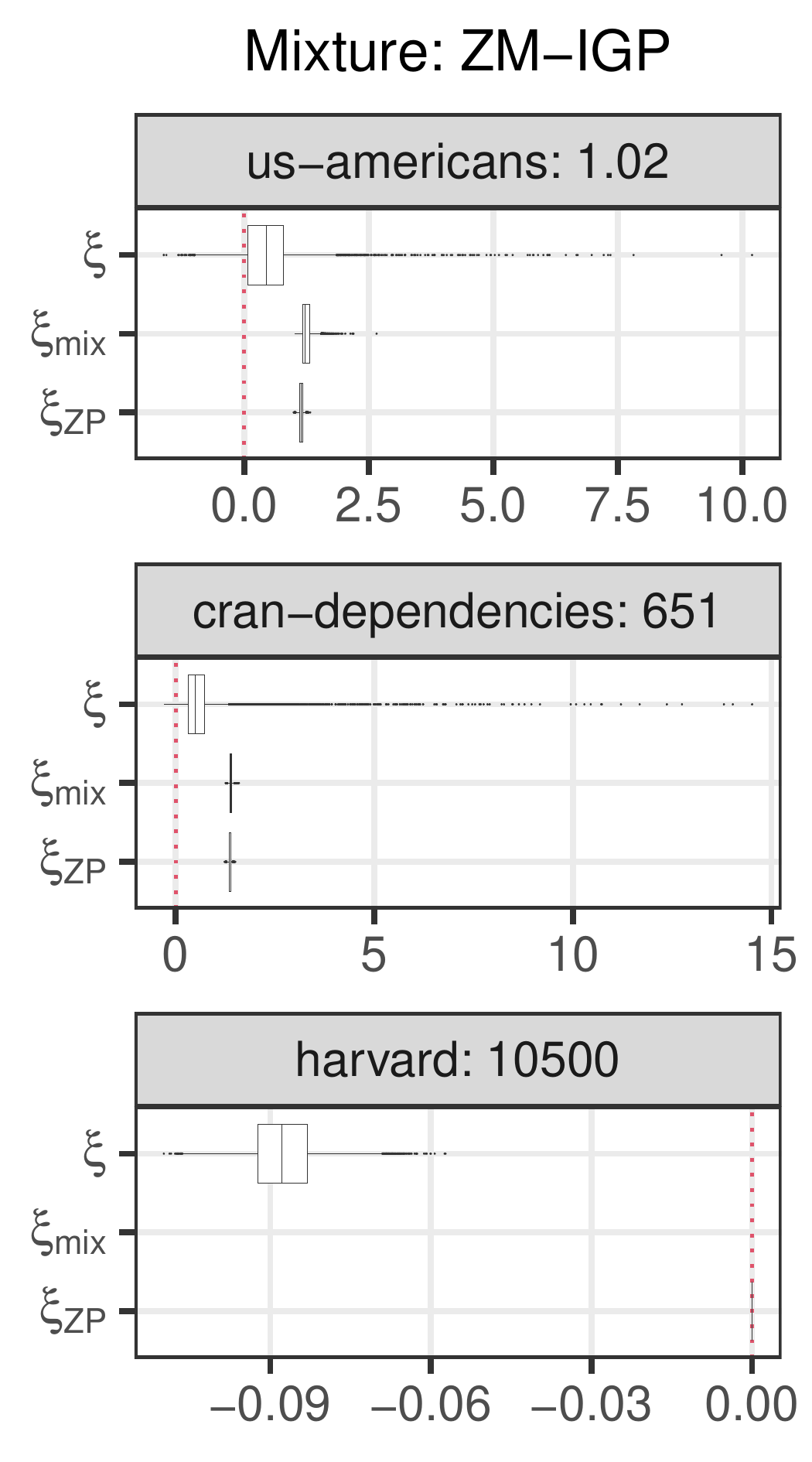} 
\includegraphics[width=0.24\linewidth]{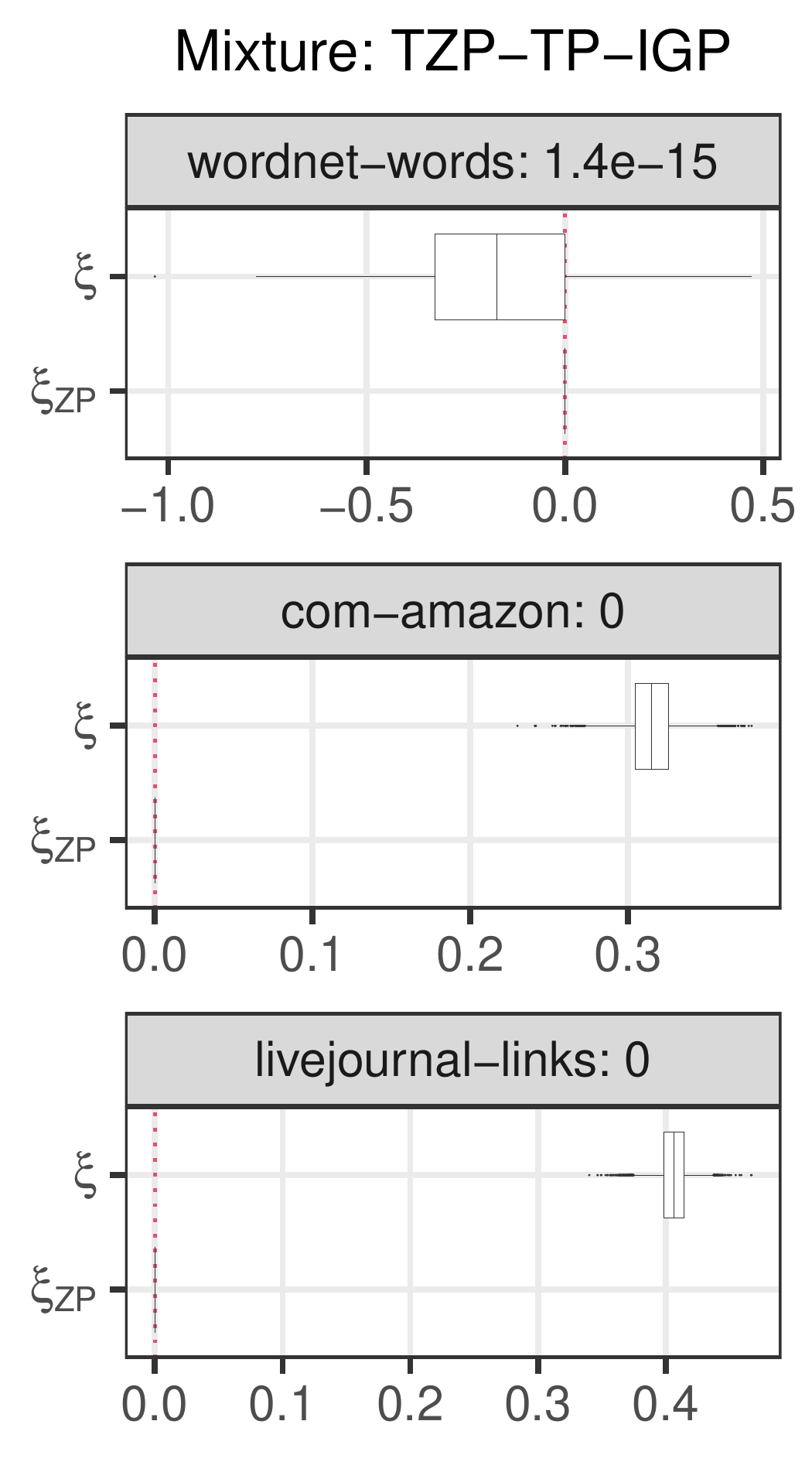} 
\includegraphics[width=0.24\linewidth]{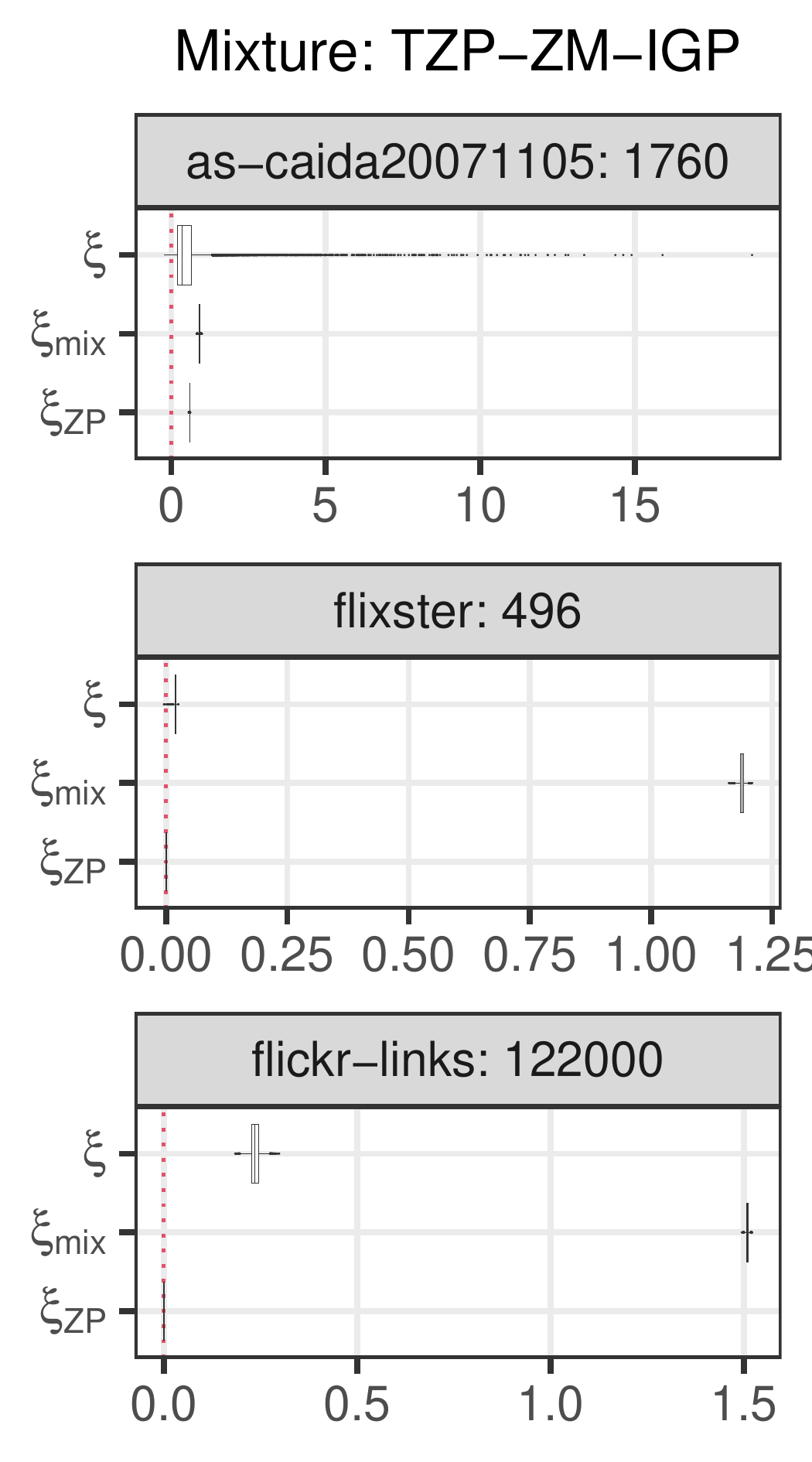} 

}

\caption{Boxplot of $\xi$ and the tail indices implied by the selected mixture and ZP distributions. The number beside the data set name is the Bayes factor (to 3 s.f.) for $\theta_{\text{mix}}=1$ relative to $\theta_{\text{mix}}\in(0,1)$. In the first and third columns, $\xi_{\text{mix}}=0$; in the other two columns, some boxes for $\xi_{\text{mix}}$ are missing as all sampled values of $\alpha_{\text{mix}}$ are smaller than 1.}\label{fig:plot-xi-alpha}
\end{figure}

\end{knitrout}

Figure \ref{fig:plot-fitted-surv} shows the empirical and model-based survival plots; the survival plot was chosen over the frequency plot as it magnifies the right tail. We only show representative data sets here, including those shown in Figure \ref{fig:plot-example-data}; the plots for all data sets are in Appendix \ref{appendix.results}. For each data set, we only show the mixture fit by either the (TZP)-ZM-IGP or (TZP)-TZP-IGP distribution based on model selection; similarly for the ZP fit by either the Zipf or polylog distribution. Due to the large sample sizes, the 95\% credible intervals of the fitted survival function by the selected mixture distribution (red) are narrow for all data sets, but they do show good fits to the whole of the data. The counterparts for the fitted ZP distribution (green) show a much poorer fit in the majority of cases. This is because the curvature of the survival function leads to the polylog distribution being selected; since this implies $\xi_{\text{ZP}}=0$, the right tail is frequently underestimated.

Figure \ref{fig:plot-fitted-surv} also gives insights into the uncertainty of $u$ and, where applicable, $v$. The threshold $u$ is usually located to the right of the region where the power law is applicable, and its posterior support is a subset of the data range, demonstrating that the inference procedure for the mixture distributions locates a division between the body and the tail regions. As expected, the threshold uncertainty tends to decrease with increasing sample size, suggesting that deviations from the body are easier to identify with larger networks.

The boxplots of $\xi$ and the implied tail indices $\xi_{\text{ZP}}$ and $\xi_{\text{mix}}$ in Figure \ref{fig:plot-xi-alpha} reinforce the need for a mixture distribution, since for most data sets, $\xi$ of the mixture distribution is non-zero while the equivalent value implied by a ZP fit ($\xi_{\text{ZP}}$) is $0$. Further, for the mixture fits in which the TP distribution is the bulk, $\xi$ is markedly different to and generally larger than $\xi_{\text{mix}}=0$. On the other hand, for cases in which the ZM distribution is the bulk, $\xi$ is generally smaller than $\xi_{\text{mix}}$. Consequently, the IGP distribution captures the right tail-heaviness that $\left(\theta_{\text{ZP}},\alpha_{\text{ZP}}\right)$ in the ZP distribution cannot. 

We revisit the four data sets introduced in Section \ref{sect.intro} to examine the performance of the mixture distributions in detail. For both ``wordnet-words'' (top row, third column) and ``flickr-links'' (bottom row, fourth column), the 3-component mixture successfully identifies the left and right tails, and provides a much better fit than the ZP distribution. The main difference is that the power law is implied in the body for ``flickr-links'' but not ``wordnet-words''; see Section \ref{sect.adequacy} for the explanation. For ``harvard'' (bottom row, second column), all sampled values of $\alpha_{\text{mix}}$ are smaller than $1$, suggesting a partial power law with an exponent that would not be possible had the Zipf distribution been fitted to the whole of the data. Relatedly, as, according to the definition in Section \ref{sect.mix2}, the implied tail index $\xi_{\text{mix}}$ is meaningful only when $\alpha_{\text{mix}}>1$, the boxplot of $\xi_{\text{mix}}$ is therefore missing in Figure \ref{fig:plot-xi-alpha} for the ``harvard'' data. Finally, for ``ucsc'' (bottom row, first column), the 2-component mixture fits the curvature in the body well, while simultaneously suggesting a tail slightly lighter than what would have been implied by fitting the ZP distribution to the whole of the data, according to the boxplots of $\xi$ and $\xi_{\text{ZP}}$ in Figure \ref{fig:plot-xi-alpha}.

\subsection{Adequacy of the power law in the body} \label{sect.adequacy}
For each data set, we ask two questions regarding the suitability of the power law. Firstly, does the power law actually hold in the body? Equivalently, is the (TZP-)ZM-IGP distribution  adequate relative to the (TZP-)TP-IGP distribution? This is answered by the Bayes factor $B_{10}$ described in Section \ref{sect.model_select}, in the label following the name of the data in Figure \ref{fig:plot-fitted-surv}. According to \cite{kr95}, $B_{10}>20$ is considered as strong evidence for $\theta_{\text{mix}}=1$ against $\theta_{\text{mix}}\in(0,1)$. Therefore, the power law can be safely said to hold for ``cran-dependencies'', ``as-caida20071105'', ``harvard'', and ``flickr-links''. On the other hand, ``moby-dick'', ``wordnet-words'' and ``com-amazon'' could be deemed to follow the power law from the log-log plot, however their Bayes factors are all effectively zero, suggesting otherwise. There is no contradiction here -- although the posterior for $\theta_{\text{mix}}$ for these four data sets is highly concentrated  close to $1$, proximity to $1$ does not translate to a high Bayes factor since $\theta_{\text{mix}}\in(0,1)$ and $\theta_{\text{mix}}=1$ imply different right tail behaviours.

Secondly, for the data sets for which the power law holds in the body, is it plausible that it holds for the whole range of values (for the 2-component mixture) or the data above $v$ (for the 3-component mixture)?  Among those identified to (partially) follow the power law, ``cran-dependencies'' and ``as-caida20071105'' are the most likely candidates since both data sets have a large amount of uncertainty of $u$ and the posterior has an upper end point close to the maximum data value. For ``as-caida20071105'', the green credible intervals in Figure \ref{fig:plot-fitted-surv} indicates a poor fit of the selected Zipf distribution to the whole of the data; fitting to the data above $v$ would result in a better fit that is closer to the mixture counterpart.

Our findings strengthen the need for the mixture distributions, and highlight the consequence of fitting the incorrect distribution. The strong evidence for the partial power law for the data sets identified in this section means that the PA model could be the network generative model \textit{up to a point}. Applying the ZP distribution with model selection to the whole of the data often leads to the polylog distribution being selected, thus overlooking the power law in the body and the potential of partial scale-free growth of the network.

\section{Discussion} \label{sect.discussion}
In this paper, we proposed a modelling and inference framework for the application of discrete extreme value mixture distributions in situations in which the power law would otherwise automatically be applied. The mixture distributions are shown to provide good fits to both the body and tail of the data for numerous examples, whilst quantifying the threshold uncertainty and without subjectively discarding most data points as in a traditional extreme value analysis. For most data sets analysed, the popular Zipf distribution is insufficient to describe the right tail of the data, which is usually lighter than would be expected from the decay of the body of the data, nor is its generalisation -- the ZP distribution -- adequate to describe the whole of the data. The mixture distribution capitalises on the usefulness of the ZP distribution by utilising its truncated version -- the TZP distribution -- for the body of the data, while capturing the right tail using the IGP distribution.

Bayesian model selection enables us to formally determine whether the power law is adequate for the body of the data fitted by the TZP distribution. It is possible to expand the model selection to between the ZP and mixture distributions, essentially testing if $u=\infty$ is adequate, or between the IGP distribution and the (2-component) mixture, essentially testing if $u=1$ is adequate. However, we deem neither model selection necessary, as it is clear from the applications that the mixture distribution outperforms the ZP distribution, and as for most data sets $u$ is comfortably away from the lower and upper end-points of the distribution, indicating the usefulness of the mixture distributions. On the other hand, if model selection between all distributions (polylog, Zipf, TP-IGP, ZM-IGP, TZP-TP-IGP, TZP-ZM-IGP) is necessary, a systematic approach would be to compare the marginal likelihood for each distribution, which can be computed within the MCMC algorithm.

As mentioned in Section \ref{sect.like}, the assumption that the degrees are independent of each other is unrealistic. However, the larger degrees, which are of more interest, are likely less dependent on each other than the smaller degrees are. Nevertheless, an avenue for research is to include such dependence in the mixture model.

Lastly and relatedly, we return to the connection between the degree distribution and network generating models. Further investigations on this are beyond the scope of this paper, however the results in Section \ref{sect.app} suggest that the PA model is not a realistic generating mechanism for many real-world examples of degree distributions. Given the larger number of networks for which the mixture distribution best describes the degree distribution, we feel that it is of great importance to investigate whether the PA or similar network generative models could be easily modified to result in the desired degree distributions. However, if the growth according to the preferential attachment model is to be studied in real networks, the two data sets identified at the end of Section \ref{sect.app} (``cran-dependencies'' and ``as-caida20071105'') would be the most suitable candidates for such an investigation.

\section*{Declaration of interest}

None.

\bibliographystyle{agsm} 

\bibliography{mixture-sn/ref_mix}

\appendix

\section*{Appendices}

\section{Proofs} \label{appendix.proofs}
\small

In this appendix, proofs are provided for results concerning
\begin{align}
  \Omega(F,n):=\left(\log\frac{S(n+1)}{S(n+2)}\right)^{-1}-\left(\log\frac{S(n)}{S(n+1)}\right)^{-1}, \label{eqn.omega}
\end{align}
where $S(n)=1-F(n)$ is the survival function of the discrete distribution concerned. The main results for the polylog and Zipf distributions are stated in Theorems \ref{theorem.polylog} and \ref{theorem.zipf}, respectively. Here, integers are denoted by $n$ instead of $x$, as the latter denotes real numbers.

\subsection{The polylog distribution} \label{appendix.polylog}
To prove that $\displaystyle\limn\Omega(F,n)=0$ for the polylog distribution, we require the following lemma:

\begin{lemma}
  If $F$ is the polylog$(\alpha,\theta,w)$ distribution, $\displaystyle\limn\frac{S(n)}{S(n+1)}=\frac{1}{\theta}$.
\end{lemma}

\begin{proof}
  For the polylog$(\alpha,\theta,w)$ distribution, for $n=w+1,w+2,\ldots$,
  \begin{align}
    f(n)&=\frac{n^{-\alpha}\theta^n}{\sum_{k=w+1}^{\infty}k^{-\alpha}\theta^k},\quad\quad\text{and}\quad\quad S(n)=\frac{\sum_{k=n+1}^{\infty}k^{-\alpha}\theta^k}{\sum_{k=w+1}^{\infty}k^{-\alpha}\theta^k},\nonumber\\
    \intertext{which implies}
    \frac{S(n+1)}{S(n)}&=\frac{\sum_{k=n+2}^{\infty}k^{-\alpha}\theta^k}{\sum_{k=n+1}^{\infty}k^{-\alpha}\theta^k}\nonumber\\
        &= \frac{(n+2)^{-\alpha}\theta^{n+2} + (n+3)^{-\alpha}\theta^{n+3} + \ldots}{(n+1)^{-\alpha}\theta^{n+1} + (n+2)^{-\alpha}\theta^{n+2} + \ldots}\nonumber\\
        &= \theta \times \frac{(n+2)^{-\alpha}\theta^{n+1} + (n+3)^{-\alpha}\theta^{n+2} + \ldots}{(n+1)^{-\alpha}\theta^{n+1} + (n+2)^{-\alpha}\theta^{n+2} + \ldots} \label{eqn.ratio}
  \end{align}
  We will consider the limit of this ratio under three cases: $\alpha=0$, $\alpha<0$, and $\alpha>0$. First, when $\alpha=0$, the fraction in Equation \ref{eqn.ratio} becomes $1$, so
  \begin{align*}
    \frac{S(n+1)}{S(n)}&=\theta \times \frac{\theta^{n+1} + \theta^{n+2} + \ldots}{\theta^{n+1} + \theta^{n+2} + \ldots} =\theta\qquad\Rightarrow\qquad\limn\frac{S(n+1)}{S(n)}=\theta
  \end{align*}
  When $\alpha<0$, $$(n+2)^{-\alpha}\theta^{n+1}>(n+1)^{-\alpha}\theta^{n+1},\qquad(n+3)^{-\alpha}\theta^{n+1}>(n+2)^{-\alpha}\theta^{n+1},$$ and so on. This means each term in the numerator in Equation \ref{eqn.ratio} is greater than its counterpart in the denominator. Subsequently, the fraction is greater than $1$, and $\displaystyle\frac{S(n+1)}{S(n)}>\theta$. On the other hand, we can continue with Equation \ref{eqn.ratio} in the following way:
  \begin{align*}
    \frac{S(n+1)}{S(n)}
    &= \theta \times \frac{(n+2)^{-\alpha}\theta^{n+1} + (n+3)^{-\alpha}\theta^{n+2} + \ldots}{(n+1)^{-\alpha}\theta^{n+1} + (n+2)^{-\alpha}\theta^{n+2} + \ldots}\\
    &= \theta\left(\frac{n+2}{n+1}\right)^{-\alpha}\times \frac{\theta^{n+1}+\left(\frac{n+3}{n+2}\right)^{-\alpha}\theta^{n+2}+\left(\frac{n+4}{n+2}\right)^{-\alpha}\theta^{n+3}+\ldots}{\theta^{n+1}+\left(\frac{n+2}{n+1}\right)^{-\alpha}\theta^{n+2}+\left(\frac{n+3}{n+1}\right)^{-\alpha}\theta^{n+3}+\ldots}\\ 
    &< \theta\left(\frac{n+2}{n+1}\right)^{-\alpha}\times \frac{\theta^{n+1}+\left(\frac{n+2}{n+1}\right)^{-\alpha}\theta^{n+2}+\left(\frac{n+3}{n+1}\right)^{-\alpha}\theta^{n+3}+\ldots}{\theta^{n+1}+\left(\frac{n+2}{n+1}\right)^{-\alpha}\theta^{n+2}+\left(\frac{n+3}{n+1}\right)^{-\alpha}\theta^{n+3}+\ldots}\\
    &= \theta\left(\frac{n+2}{n+1}\right)^{-\alpha}.
  \end{align*}
  Combining the inequalities and using the squeeze theorem, we have
  \begin{eqnarray*}
    \displaystyle\theta<&\displaystyle\frac{S(n+1)}{S(n)}&<\theta\left(\frac{n+2}{n+1}\right)^{-\alpha}\\
    \limn\theta\leq&\displaystyle\limn\frac{S(n+1)}{S(n)}&\leq\limn\theta\left(\frac{n+2}{n+1}\right)^{-\alpha}\\
    \theta\leq&\displaystyle\limn\frac{S(n+1)}{S(n)}&\leq\theta\\
   &\displaystyle\limn\frac{S(n+1)}{S(n)}&=\theta 
  \end{eqnarray*}
  Lastly, when $\alpha>0$, the derivations are the same as when $\alpha<0$, but with all inequality signs reversed. Combining all three cases, and as $\theta\in(0,1)$ implies that $1/\theta$ is a finite positive number,
  \begin{align*}
    \limn\frac{S(n+1)}{S(n)}=\theta\qquad\Rightarrow\qquad\limn\frac{S(n)}{S(n+1)}=\frac{1}{\theta}.
  \end{align*}
\end{proof}

We now state the main result:
\begin{theorem}
  If $F$ is the polylog$(\alpha,\theta,w)$ distribution, $\displaystyle\limn\Omega(F,n)=0$.
  \label{theorem.polylog}
\end{theorem}

\begin{proof}
  As the polylog$(\alpha,\theta,w)$ distribution is concerned, we have the following result for $\theta$:
  \begin{align*}
    \theta\in(0,1)\qquad\Rightarrow\qquad\frac{1}{\theta}>1\qquad\Rightarrow\qquad\log\frac{1}{\theta}>0\qquad\Rightarrow\qquad\left(\log\frac{1}{\theta}\right)^{-1}>0.
  \end{align*}
  Now, since $\displaystyle\frac{S(n+1)}{S(n+2)}$ has the same limit as $\displaystyle\frac{S(n)}{S(n+1)}$ as $n\rightarrow\infty$, we complete the proof as 
  \begin{align*}
    \limn\Omega(F,n)&=\limn\left[\left(\log\frac{S(n+1)}{S(n+2)}\right)^{-1}-\left(\log\frac{S(n)}{S(n+1)}\right)^{-1}\right]\\
                    &=\left(\log\limn\frac{S(n+1)}{S(n+2)}\right)^{-1}-\left(\log\limn\frac{S(n)}{S(n+1)}\right)^{-1}\\
                    &=\left(\log\frac{1}{\theta}\right)^{-1}-\left(\log\frac{1}{\theta}\right)^{-1}=0.
  \end{align*}
\end{proof}

\subsection{The Zipf distribution} \label{appendix.zipf}
In this section, we prove several lemmas and lastly $\displaystyle\limn\Omega(F,n)=\frac{1}{\alpha-1}$ for the Zipf distribution with exponent $\alpha$. Unless otherwise specified, $x$ and $n$ refers to positive real numbers and positive integers, respectively. Note that the proofs for the polylog distribution (where $\theta\in(0,1)$) cannot be applied here for the Zipf distribution (where $\theta=1$), as $\displaystyle\left(\log\frac{1}{\theta}\right)^{-1}$ is now undefined.

\begin{lemma}
  $$ \limx{}(x+1)^{\alpha-1}\zeta(\alpha,x+2)=\frac{1}{\alpha-1}. $$
  \label{lemma.1overalphaminus1}
\end{lemma}

\begin{proof}
  Equation (2.3) of \cite{berndt72} asserts that, for $\alpha>1$, 
  \begin{align*}
    \zeta(\alpha,x)&=x^{-\alpha}+\frac{x^{1-\alpha}}{\alpha-1}-\alpha\int_0^{\infty}(u-\lfloor{}u\rfloor)(u+x)^{-\alpha-1}du\\
    &=\frac{1}{x^{\alpha}}+\frac{1}{(\alpha-1)x^{\alpha-1}}-\alpha\int_0^{\infty}\frac{u-\lfloor{}u\rfloor}{(u+x)^{\alpha+1}}du.
  \end{align*}
  Replacing $x$ by $x+2$ and multiplying both sides by $(x+1)^{\alpha-1}$,
  \begin{align*}
    \limx(x+1)^{\alpha-1}\zeta(\alpha,x+2)&=\limx\frac{(x+1)^{\alpha-1}}{(x+2)^{\alpha}}+\limx\frac{(x+1)^{\alpha-1}}{(\alpha-1)(x+2)^{\alpha-1}}\\
    &\quad-\alpha\int_0^{\infty}\limx\frac{(u-\lfloor{}u\rfloor)(x+1)^{\alpha-1}}{(u+x+2)^{\alpha+1}}du\\
    &=\limx\left(\frac{x+1}{x+2}\right)^{\alpha-1}\times\limx\left(\frac{1}{x+1}\right)+\frac{1}{\alpha-1}\limx\left(\frac{x+1}{x+2}\right)^{\alpha-1}\\
    &\quad-\alpha\int_0^{\infty}\left[(u-\lfloor{}u\rfloor)\limx\left(\frac{x+1}{u+x+2}\right)^{\alpha-1}\limx\frac{1}{(u+x+2)^2}\right]du\\
    &=1\times0+\frac{1}{\alpha-1}\times1-\alpha\int_0^{\infty}\left[(u-\lfloor{}u\rfloor)\times1\times0\right]du\\
    &=\frac{1}{\alpha-1}.
  \end{align*}
\end{proof}


\begin{lemma}
  $$ \limx\left[(x+2)^{\alpha}\exp\left(-\frac{\alpha-1}{x+2}\right)-(x+1)^{\alpha}\right]\zeta(\alpha,x+2)=\frac{1}{\alpha-1}. $$
  \label{lemma.lowerbound}
\end{lemma}

\begin{proof}
  \begin{align*}
    \limx\frac{(x+2)^{\alpha}\exp\left(-\frac{\alpha-1}{x+2}\right)-(x+1)^{\alpha}}{(x+1)^{\alpha-1}}&=\limx\left[(x+2)\left(\frac{x+2}{x+1}\right)^{\alpha-1}\exp\left(-\frac{\alpha+1}{x+2}\right)-(x+1)\right]\\
    &=\limx(x+2)\left[\left(\frac{x+2}{x+1}\right)^{\alpha-1}\exp\left(-\frac{\alpha+1}{x+2}\right)-1\right]+1\\
    &=1+\limx\frac{\left(1+\frac{1}{x+1}\right)^{\alpha-1}\exp\left(-\frac{\alpha+1}{x+2}\right)-1}{\frac{1}{x+2}}\\
    \intertext{(Using L'Hopital's rule)}
    &=1+\limx\left[\frac{(\alpha-1)\left(\frac{x+2}{x+1}\right)^{\alpha-2}\left(-\frac{1}{(x+1)^2}\right)\exp\left(-\frac{\alpha+1}{x+2}\right)}{-\left(\frac{1}{(x+2)^2}\right)}\right.\\
    &\qquad\qquad\qquad+\left.\frac{\left(1+\frac{1}{x+1}\right)^{\alpha-1}\exp\left(-\frac{\alpha+1}{x+2}\right)\left(\frac{\alpha+1}{(x+2)^2}\right)}{-\left(\frac{1}{(x+2)^2}\right)}\right]\\
    &=1+\limx\left[(\alpha-1)\left(\frac{x+2}{x+1}\right)^{\alpha}\exp\left(-\frac{\alpha+1}{x+2}\right)\right.\\
    &\qquad\qquad\qquad-\left.\left(1+\frac{1}{x+1}\right)^{\alpha-1}\exp\left(-\frac{\alpha+1}{x+2}\right)(\alpha-1)\right]\\
    &=1+(\alpha-1)\times1\times1-1\times1\times(\alpha-1)=1.
  \end{align*}
  Therefore, using Lemma \ref{lemma.1overalphaminus1},
  \begin{align*}
    \limx\left[(x+2)^{\alpha}\exp\left(-\frac{\alpha-1}{x+2}\right)-(x+1)^{\alpha}\right]\zeta(\alpha,x+2)&=\limx\left[\frac{(x+2)^{\alpha}\exp\left(-\frac{\alpha-1}{x+2}\right)-(x+1)^{\alpha}}{(x+1)^{\alpha-1}}\right.\\
    &\qquad\qquad\times \left.(x+1)^{\alpha-1}\zeta(\alpha,x+2)\right]\\
    &=1\times\frac{1}{\alpha-1}=\frac{1}{\alpha-1}.
  \end{align*}
\end{proof}

\begin{lemma}
  $$ \limx \frac{(x+2)^{\alpha}\left(\frac{x+2}{x+3}\right)^{\alpha-1}-(x+1)^{\alpha}}{(x+2)^{\alpha}-(x+1)^{\alpha}}=\frac{1}{\alpha}. $$
  \label{lemma.upperbound}
\end{lemma}

\begin{proof}
  \begin{align*}
    \limx \frac{(x+2)^{\alpha}\left(\frac{x+2}{x+3}\right)^{\alpha-1}-(x+1)^{\alpha}}{(x+2)^{\alpha}-(x+1)^{\alpha}}&=\limx \frac{\left(\frac{x+2}{x+3}\right)^{\alpha-1}-\left(\frac{x+1}{x+2}\right)^{\alpha}}{1-\left(\frac{x+1}{x+2}\right)^{\alpha}}\\
    &=1+\limx \frac{\left(\frac{x+2}{x+3}\right)^{\alpha-1}-1}{1-\left(\frac{x+1}{x+2}\right)^{\alpha}}\\
    &=1+\limx \frac{\left(1-\frac{1}{x+3}\right)^{\alpha-1}-1}{1-\left(1-\frac{1}{x+2}\right)^{\alpha}}\\
    \intertext{(Using L'Hopital's rule)}
    {}&=1+\limx \frac{(\alpha-1)\left(1-\frac{1}{x+3}\right)^{\alpha-2}\left(\frac{1}{x+3}\right)^2}{-\alpha\left(1-\frac{1}{x+2}\right)^{\alpha-1}\left(\frac{1}{x+2}\right)^2}\\
    &=1-\frac{\alpha-1}{\alpha}\limx\frac{\left(\frac{x+2}{x+3}\right)^{\alpha-2}}{\left(\frac{x+1}{x+2}\right)^{\alpha-1}}\limx\left(\frac{x+2}{x+3}\right)^2\\
    &=1-\frac{\alpha-1}{\alpha}\times1\times1 = \frac{1}{\alpha}.
  \end{align*}
\end{proof}

\begin{lemma}
  $$ \limx \left[(x+2)^{\alpha}\zeta(\alpha,x+3)-(x+1)^{\alpha}\zeta(\alpha,x+2)\right]=\frac{1}{\alpha-1}. $$
  \label{lemma.difference}
\end{lemma}

\begin{proof}
  According to \cite{alzer05}, for $x>0$ and $\alpha>1$,
  \begin{align*}
    \exp\left(-\frac{\alpha-1}{x+2}\right)<\frac{\zeta(\alpha,x+3)}{\zeta(\alpha,x+2)}<\left(\frac{x+2}{x+3}\right)^{\alpha-1}.
  \end{align*}
  Multiplying all sides by $(x+2)^{\alpha}$, subtracting $(x+1)^{\alpha}$, and multiplying by $\zeta(\alpha,x+2)$,
  \begin{align*}
    &\left[(x+2)^{\alpha}\exp\left(-\frac{\alpha-1}{x+2}\right)-(x+1)^{\alpha}\right]\zeta(\alpha,x+2)\\
    <~&(x+2)^{\alpha}\zeta(\alpha,x+3)-(x+1)^{\alpha}\zeta(\alpha,x+2)\\
    <~&\left[(x+2)^{\alpha}\left(\frac{x+2}{x+3}\right)^{\alpha-1}-(x+1)^{\alpha}\right]\zeta(\alpha,x+2)
  \end{align*}
  \cite{cl84} showed, in proving their Lemma 3, that $$\frac{\alpha-1}{\alpha}\zeta(\alpha,x+2)-[(x+2)^{\alpha}-(x+1)^{\alpha}]^{-1}$$ is a monotonic function of $x$ and increasing to $0$ as $\xinfty$. Equivalently, $$\zeta(\alpha,x+2)<\left[(x+2)^{\alpha}-(x+1)^{\alpha}\right]^{-1}\frac{\alpha}{\alpha-1}.$$ Therefore, we can change the inequality to
  \begin{align*}
    &\left[(x+2)^{\alpha}\exp\left(-\frac{\alpha-1}{x+2}\right)-(x+1)^{\alpha}\right]\zeta(\alpha,x+2)\\
    <~&(x+2)^{\alpha}\zeta(\alpha,x+3)-(x+1)^{\alpha}\zeta(\alpha,x+2)\\
    <~&\frac{(x+2)^{\alpha}\left(\frac{x+2}{x+3}\right)^{\alpha-1}-(x+1)^{\alpha}}{(x+2)^{\alpha}-(x+1)^{\alpha}}\times\frac{\alpha}{\alpha-1}
  \end{align*}
  From Lemma \ref{lemma.lowerbound}, the lower bound in the first line converges to $\displaystyle\frac{1}{\alpha-1}$ as $\xinfty$. From Lemma \ref{lemma.upperbound}, the upper bound in the last line converges to $\displaystyle\frac{\alpha}{\alpha-1}\times\frac{1}{\alpha}=\frac{1}{\alpha-1}$ i.e. the same limit as $\xinfty$. Therefore, using the squeeze theorem, $$\limx\left[(x+2)^{\alpha}\zeta(\alpha,x+3)-(x+1)^{\alpha}\zeta(\alpha,x+2)\right]=\frac{1}{\alpha-1}.$$
\end{proof}

\begin{lemma}
  For the Zipf$(\alpha,w)$ distribution, $$\limn \frac{f(n+1)}{S(n+1)}=0,\quad\text{and}\quad\limn\left(\log\frac{S(n)}{S(n+1)}\right)^{-1}-\left(\frac{f(n+1)}{S(n+1)}\right)^{-1}=\frac{1}{2}.$$
  \label{lemma.penultimate}
\end{lemma}
  
\begin{proof}
  Lemma \ref{lemma.1overalphaminus1} applies to real $x$, and thus integer $n$. Therefore, $$\limn (n+1)^{\alpha-1}\zeta(\alpha,n+2)=\frac{1}{\alpha-1}\quad\Rightarrow\quad\limn\frac{1}{(n+1)^{\alpha-1}\zeta(\alpha,n+2)}=\alpha-1.$$ For the Zipf$(\alpha,w)$ distribution, $$f(n)=\frac{n^{-\alpha}}{\zeta(\alpha,w+1)}\qquad\text{and}\qquad S(n)=\frac{\zeta(\alpha,n+1)}{\zeta(\alpha,w+1)}.$$ Substituting $n$ by $n+1$, dividing the former by the latter, and taking the limit as $n\rightarrow\infty$,
  \begin{align*}
    \limn\frac{f(n+1)}{S(n+1)}&=\limn\left[\frac{(n+1)^{-\alpha}}{\zeta(\alpha,n+2)}\right]\\
    &=\limn\frac{1}{n+1}\times\frac{1}{(n+1)^{\alpha-1}\zeta(\alpha,n+2)}\\
    &=\limn\frac{1}{n+1}\times\limn\frac{1}{(n+1)^{\alpha-1}\zeta(\alpha,n+2)}\\
    &=0\times(\alpha-1)=0.
  \end{align*}
  Using the fact $$\lim_{z\rightarrow0}\left[\log(1+z)\right]^{-1}-z^{-1}=\frac{1}{2},$$ which has been used by \cite{shimura12} and can be proved by using the L'Hopital's rule, and substituting $z$ by $\displaystyle\frac{f(n+1)}{S(n+1)}$, which goes to $0$ as $n\rightarrow\infty$, we have
  \begin{align*}
    \limn\left[\log\left(1+\frac{f(n+1)}{S(n+1)}\right)\right]^{-1}-\left(\frac{f(n+1)}{S(n+1)}\right)^{-1}=\frac{1}{2}.
  \end{align*}
  As $S(n)=S(n+1)+f(n+1)$, $$1+\frac{f(n+1)}{S(n+1)}=\frac{S(n)}{S(n+1)},$$
  hence the second statement of the lemma.
\end{proof}

\begin{theorem}
  For the Zipf$(\alpha,w)$ distribution, $$ \limn\Omega(F,n)=\frac{1}{\alpha-1}. $$
  \label{theorem.zipf}
\end{theorem}

\begin{proof}
  \begin{align*}
    \Omega(F,n)&=\left(\log\frac{S(n+1)}{S(n+2)}\right)^{-1}-\left(\log\frac{S(n)}{S(n+1)}\right)^{-1}\\
    &=\left[\left(\log\frac{S(n+1)}{S(n+2)}\right)^{-1}-\left(\frac{f(n+2)}{S(n+2)}\right)^{-1}\right]-\left[\left(\log\frac{S(n)}{S(n+1)}\right)^{-1}-\left(\frac{f(n+1)}{S(n+1)}\right)^{-1}\right]\\
    &\quad+\left[\left(\frac{f(n+2)}{S(n+2)}\right)^{-1}-\left(\frac{f(n+1)}{S(n+1)}\right)^{-1}\right]\\
    &=\left[\left(\log\frac{S(n+1)}{S(n+2)}\right)^{-1}-\left(\frac{f(n+2)}{S(n+2)}\right)^{-1}\right]-\left[\left(\log\frac{S(n)}{S(n+1)}\right)^{-1}-\left(\frac{f(n+1)}{S(n+1)}\right)^{-1}\right]\\
    &\quad+\left[(n+2)^{\alpha}\zeta(\alpha,n+3)-(n+1)^{\alpha}\zeta(\alpha,n+2)\right].
  \end{align*}
  Lemma \ref{lemma.difference} implies that the limit of the last line is $\displaystyle\frac{1}{\alpha-1}$. Therefore, combined with Lemma \ref{lemma.penultimate},
  \begin{align*}
    \limn\Omega(F,n)&=\limn\left[\left(\log\frac{S(n+1)}{S(n+2)}\right)^{-1}-\left(\frac{f(n+2)}{S(n+2)}\right)^{-1}\right]\\
    &\quad-\limn\left[\left(\log\frac{S(n)}{S(n+1)}\right)^{-1}-\left(\frac{f(n+1)}{S(n+1)}\right)^{-1}\right]\\
    &\quad+\limn\left[(n+2)^{\alpha}\zeta(\alpha,n+3)-(n+1)^{\alpha}\zeta(\alpha,n+2)\right]\\
    &=\frac{1}{2}-\frac{1}{2}+\frac{1}{\alpha-1}=\frac{1}{\alpha-1}.
  \end{align*}
\end{proof}

\section{Inference algorithm} \label{appendix.inference}
Details of the inference procedure are provided in this appendix, including obtaining the set of $u$'s (and $v$'s) using the profile log-likelihood, and the steps of the Markov chain Monte Carlo (MCMC) algorithm. The 2-component mixture is used throughout for illustration; steps for the 3-component mixture are similar.

\subsection{Obtaining the set of $u$ via the profile log-likelihood} \label{appendix.profile}
As will be seen in Section \ref{appendix.mcmc}, when $u$ is sampled, the conditional posterior has to be computed for each value of a finite set. A naive starting point of this finite set is all integers within the data range. However, the size of this set is linear with the computational cost, and in practice only a handful of values will ever be sampled. Therefore, it is useful to trim the finite set, from all integers within the data range, down to the ``most probable'' values. We decide the most probable values to be those with profile log-likelihoods that are at least the maximum log-likelihood minus $11$. This is because a difference of $11$ on the log scale translates to $\exp(11)\approx\ensuremath{6\times 10^{4}}$ on the original scale, and we deem a distribution that includes thresholds with a density $1/(\ensuremath{6\times 10^{4}})$ of the maximum a sufficiently good approximation of the posterior distribution for $u$. 

To illustrate this procedure, Figure \ref{fig:appendix-moby} shows the profile log-likelihood of $u$, for one of the data sets (``moby-dick'') reported in the main paper. The $u$'s that corresponds to a value of the profile log-likelihood between the two red dashed lines, which are $11$ apart, will be included in the set of most probable thresholds for step \ref{item.u} in Section \ref{appendix.mcmc}.

\begin{knitrout}
\definecolor{shadecolor}{rgb}{0.969, 0.969, 0.969}\color{fgcolor}\begin{figure}[!h]

{\centering \includegraphics[width=0.8\linewidth]{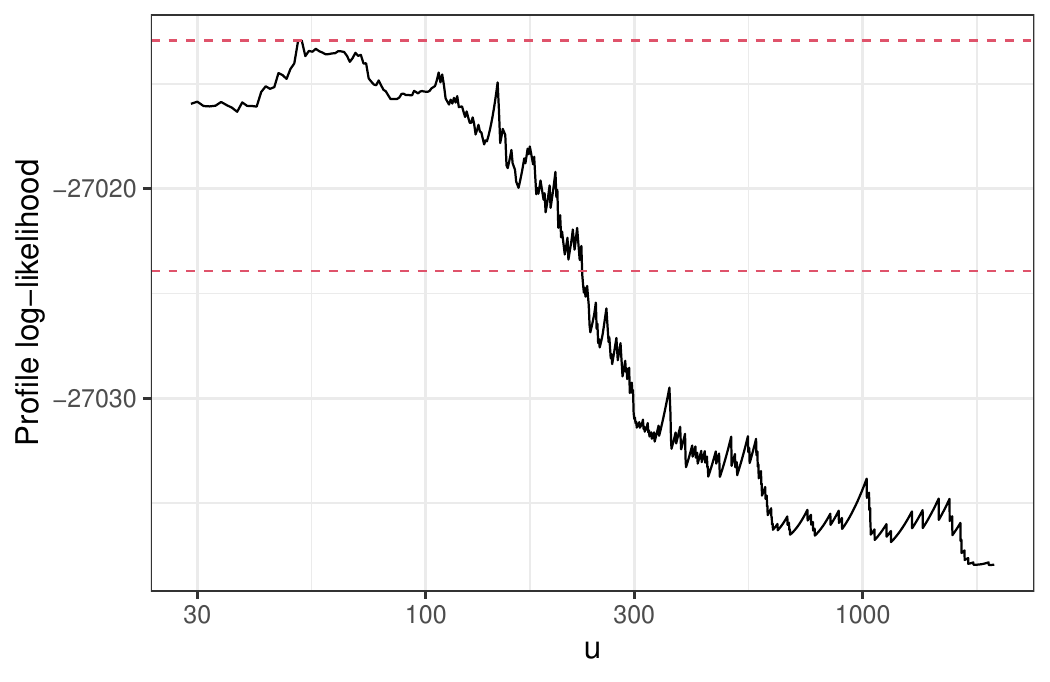} 

}

\caption[Profile log-likelihood for the Moby Dick data]{Profile log-likelihood for the Moby Dick data.}\label{fig:appendix-moby}
\end{figure}

\end{knitrout}

The procedure is similar for the 3-component mixture, but across the combinations of the two thresholds. For each $(v,u)$, $(\alpha_1,\theta_1)$, $(\alpha_{\text{mix}},\theta_{\text{mix}})$ and $(\sigma,\xi)$ are optimised separately, and the combinations of $v$ and $u$ with the required profile log-likelihood become the finite set of most probable values. Subsequently, the 3-component mixture takes longer than the 2-component mixture for the same data set, but much shorter than without this selection procedure using the profile log-likelihood.

\subsection{Markov chain Monte Carlo} \label{appendix.mcmc}
A Metropolis-within-Gibbs algorithm is implemented here for the mixture distributions. The essence is that one or more parameters are sampled from their conditional distribution given the other parameters. It is natural that $\alpha_{\text{mix}}$ and $\theta_{\text{mix}}$ are more correlated with each other than with the remaining parameters, and so they will be jointly sampled. This applies to $\sigma$ and $\xi$ too. The threshold $u$ will be sampled from the set of most probable values described above. Lastly, the model choice $M$ will be sampled by a Gibbs step. To avoid confusion that arises from using the same symbol for the variable and the value, we use $\gamma:=\Pr(M=1)$ and $1-\gamma$ for the prior model probabilities. The steps of the algorithm are as follows:

\begin{enumerate}
  \item \label{item.current} The current values in the chain are $\alpha_{\text{mix}}$, $\theta_{\text{mix}}$, $u$, $\sigma$, $\xi$ and $M$. It is implied that $\theta_{\text{mix}}=1\Leftrightarrow{}M=1$, and $\theta_{\text{mix}}\in(0,1)\Leftrightarrow{}M=0$. The steps below will preserve this relationship.
  \item \label{item.alpha_theta} 
    \begin{enumerate}
    \item If $M=0$, propose $\alpha_{\text{mix}}^{*}$ and $\theta_{\text{mix}}^{*}$ from a symmetrical density $q(\cdot|\alpha_{\text{mix}},\theta_{\text{mix}})$ and accept $(\alpha_{\text{mix}}^{*},\theta_{\text{mix}}^{*})$ with probability \footnotesize$$\min\left(1,\frac{\pi(\alpha_{\text{mix}}^{*},\theta_{\text{mix}}^{*},u,\sigma,\xi,M|\bsx,\bsc)}{\pi(\alpha_{\text{mix}},\theta_{\text{mix}},u,\sigma,\xi,M|\bsx,\bsc)}\right) = \min\left(1,\frac{L_2(\alpha_{\text{mix}}^{*},\theta_{\text{mix}}^{*},u,\sigma,\xi|\bsx,\bsc)\pi(\alpha_{\text{mix}}^{*})\pi(\theta_{\text{mix}}^{*})}{L_2(\alpha_{\text{mix}},\theta_{\text{mix}},u,\sigma,\xi|\bsx,\bsc)\pi(\alpha_{\text{mix}})\pi(\theta_{\text{mix}})}\right) $$\normalsize as the new current value of $(\alpha_{\text{mix}},\theta_{\text{mix}})$, where $\pi(\theta_{\text{mix}})$ is $\text{Beta}(1,1)$.
    \item If $M=1$, propose $\alpha_{\text{mix}}^{*}$ from a symmetrical density $q(\cdot|\alpha_{\text{mix}})$ and accept $(\alpha_{\text{mix}}^{*})$ with probability \footnotesize$$\min\left(1,\frac{\pi(\alpha_{\text{mix}}^{*},1,u,\sigma,\xi,M|\bsx,\bsc)}{\pi(\alpha_{\text{mix}},1,u,\sigma,\xi,M|\bsx,\bsc)}\right) = \min\left(1,\frac{L_2(\alpha_{\text{mix}}^{*},1,u,\sigma,\xi|\bsx,\bsc)\pi(\alpha_{\text{mix}}^{*})}{L_2(\alpha_{\text{mix}},1,u,\sigma,\xi|\bsx,\bsc)\pi(\alpha_{\text{mix}})}\right) $$\normalsize as the new current value of $\alpha_{\text{mix}}$.
    \end{enumerate}
  \item \label{item.sigma_xi} Propose $\sigma^{*}$ and $\xi^{*}$ from a symmetrical density $q(\cdot|\sigma,\xi)$ and accept $(\sigma^{*},\xi^{*})$ with probability \footnotesize$$\min\left(1,\frac{\pi(\alpha_{\text{mix}},\theta_{\text{mix}},u,\sigma^{*},\xi^{*},M|\bsx,\bsc)}{\pi(\alpha_{\text{mix}},\theta_{\text{mix}},u,\sigma,\xi,M|\bsx,\bsc)}\right) = \min\left(1,\frac{L_2(\alpha_{\text{mix}},\theta_{\text{mix}},u,\sigma^{*},\xi^{*}|\bsx,\bsc)\pi(\sigma^{*})\pi(\xi^{*})}{L_2(\alpha_{\text{mix}},\theta_{\text{mix}},u,\sigma,\xi|\bsx,\bsc)\pi(\sigma)\pi(\xi)}\right)$$\normalsize as the new current value of $(\sigma,\xi)$.
  \item \label{item.u} Denote the finite set of most probable thresholds, described in the main paper, by $\{u_1,u_2,\ldots,u_K\}$. For $k=1,2,\ldots,K$, calculate \footnotesize$$L_2(\alpha_{\text{mix}},\theta_{\text{mix}},u_k,\sigma,\xi|\bsx,\bsc)\pi(\psi_{u_k}),$$\normalsize where $\psi_{u_k}=\sum_{i=1}^m\mathbb{I}_{\{x_i>u_k\}}/m$, the empirical unique exceedance rate of $u_k$. Sample one threshold from $\{u_1,u_2,\ldots,u_K\}$, with probabilities \footnotesize$$\left\{\frac{L_2(\alpha_{\text{mix}},\theta_{\text{mix}},u_1,\sigma,\xi|\bsx,\bsc)\pi(\psi_{u_1})}{\sum_{k=1}^K L_2(\alpha_{\text{mix}},\theta_{\text{mix}},u_k,\sigma,\xi|\bsx,\bsc)\pi(\psi_{u_k})},\ldots,\frac{L_2(\alpha_{\text{mix}},\theta_{\text{mix}},u_K,\sigma,\xi|\bsx,\bsc)\pi(\psi_{u_K})}{\sum_{k=1}^K L_2(\alpha_{\text{mix}},\theta_{\text{mix}},u_k,\sigma,\xi|\bsx,\bsc)\pi(\psi_{u_k})}\right\},$$\normalsize and set that to the current value of $u$.
  \item \label{item.pseudo} 
    \begin{enumerate}
    \item If $M=0$, set $\theta_{\text{mix}}^{'}=\theta_{\text{mix}}$.
    \item If $M=1$, draw a value from $\tilde{\pi}(\theta_{\text{mix}})$, denoted by $\theta_{\text{mix}}^{'}$, where $\tilde{\pi}(\theta_{\text{mix}})$ is a \textit{pseudoprior}, the choice of which does not affect the posterior for $\bseta_2$ and $M$, as long as it has the same support as $\pi(\theta_{\text{mix}})$ (in this case $(0,1)$),  but may affect the efficiency of the MCMC algorithm.
    \end{enumerate}
  \item \label{item.M} Set the new current values of $(M,\theta_{\text{mix}})$ to $(0,\theta_{\text{mix}}^{'})$ and $(1,1)$ with probabilities \small$$ \begin{aligned} \pi(M=0|\alpha_{\text{mix}},\theta_{\text{mix}},u,\sigma,\xi,\bsx,\bsc)&\propto{}L_2(\alpha_{\text{mix}},\theta_{\text{mix}}^{'},u,\sigma,\xi|\bsx,\bsc)\pi(\theta_{\text{mix}}^{'})\times(1-\gamma),~~\text{and}\\ \pi(M=1|\alpha_{\text{mix}},\theta_{\text{mix}},u,\sigma,\xi,\bsx,\bsc)&\propto{}L_2(\alpha_{\text{mix}},1,u,\sigma,\xi|\bsx,\bsc)\tilde{\pi}(\theta_{\text{mix}}^{'})\times{}\gamma,\end{aligned}$$\normalsize respectively. Note the presence of the pseudoprior $\tilde{\pi}$ in the second line.
  \item \label{item.repeat} Repeat steps \ref{item.current} to \ref{item.M} until the algorithm has converged.
\end{enumerate}

\section{Full results in application} \label{appendix.results}

In this appendix, we show the results of applying the ZP and mixture distributions to all data sets mentioned in Section 5 of the paper. Figure \ref{fig:appendix-plot-fitted-surv} shows the empirical survival plot overlaid by the model-based counterparts with credible intervals, and is the full version of Figure 4 in the paper. Figure \ref{fig:appendix-plot-xi-alpha} shows the boxplots of $\xi$ and the implied tail indices $\xi_{\text{ZP}}$ and $\xi_{\text{mix}}$, and is the full version of Figure 5 in the paper.

\begin{knitrout}
\definecolor{shadecolor}{rgb}{0.969, 0.969, 0.969}\color{fgcolor}\begin{figure}

{\centering \includegraphics[width=0.24\linewidth]{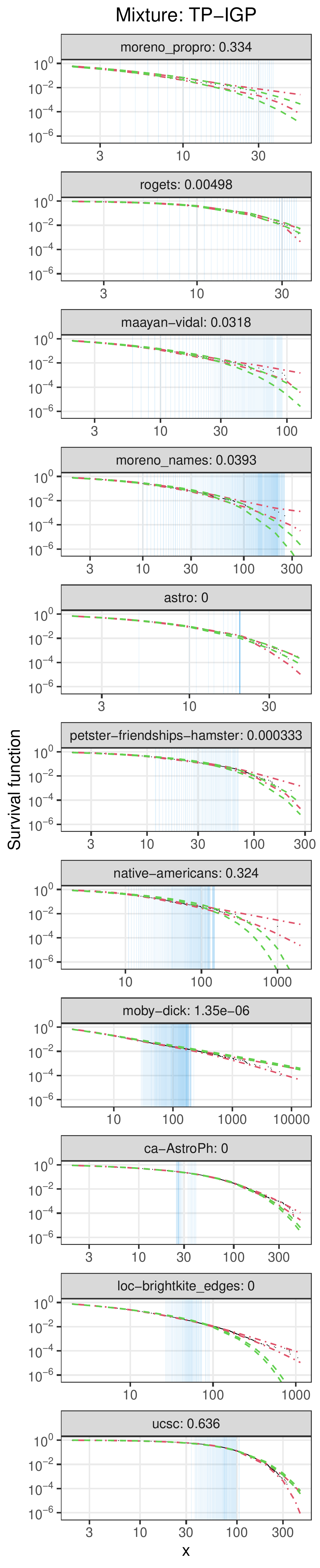} 
\includegraphics[width=0.24\linewidth]{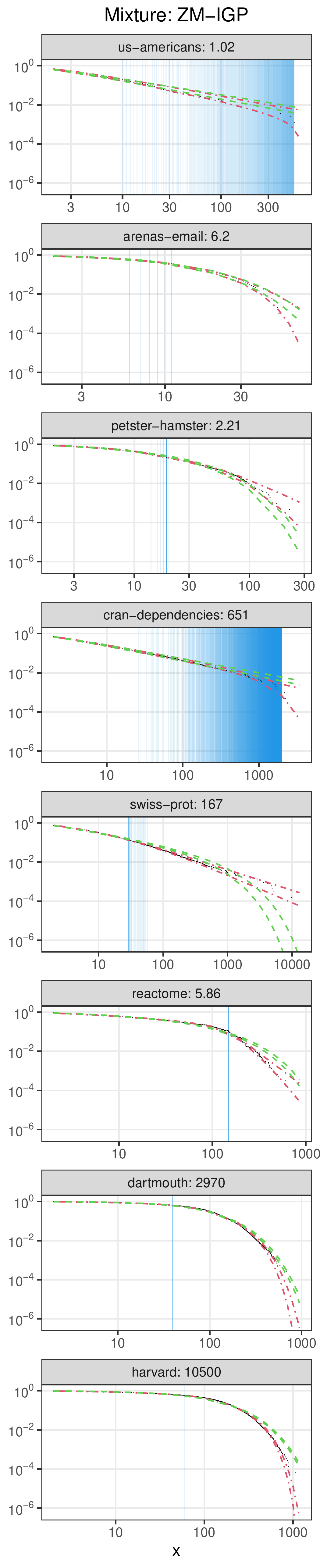} 
\includegraphics[width=0.24\linewidth]{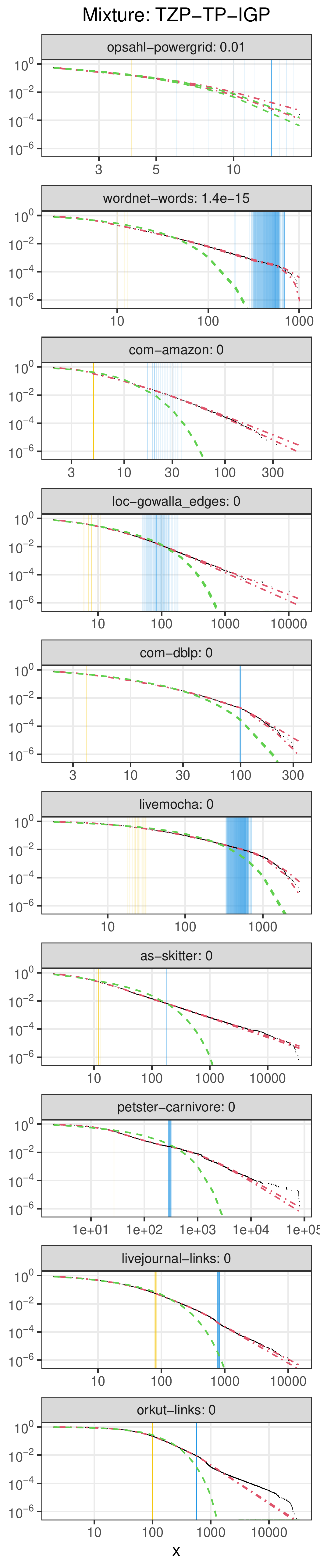} 
\includegraphics[width=0.24\linewidth]{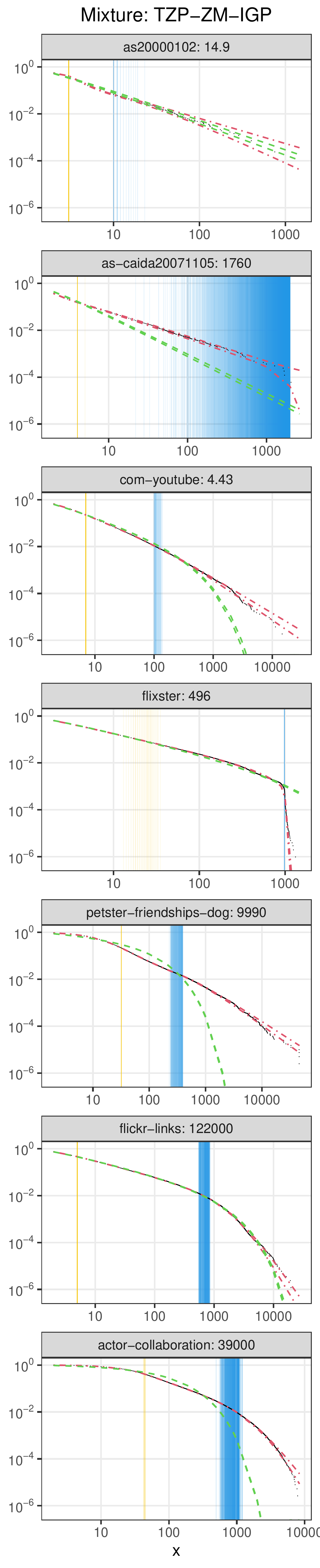} 

}

\caption[Survival function with credible intervals for the selected mixture (red) and ZP (green) distributions]{Survival function with credible intervals for the selected mixture (red) and ZP (green) distributions. The blue and yellow bands represent the posterior of $u$ and $v$, respectively, for the mixture distribution fits. The number beside the data set name is the Bayes factor (to 3 s.f.) for $\theta_{\text{mix}}=1$ relative to $\theta_{\text{mix}}\in(0,1)$.}\label{fig:appendix-plot-fitted-surv}
\end{figure}

\end{knitrout}

\begin{knitrout}
\definecolor{shadecolor}{rgb}{0.969, 0.969, 0.969}\color{fgcolor}\begin{figure}

{\centering \includegraphics[width=0.24\linewidth]{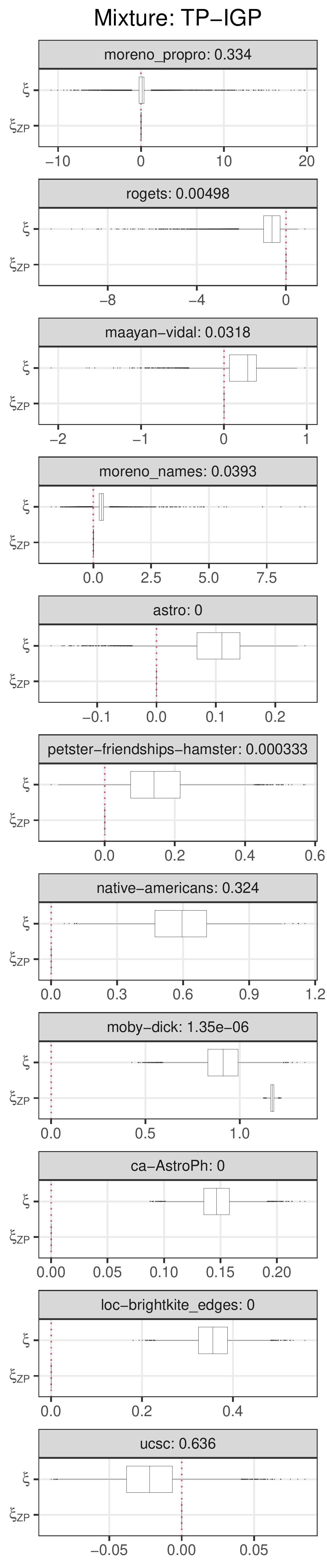} 
\includegraphics[width=0.24\linewidth]{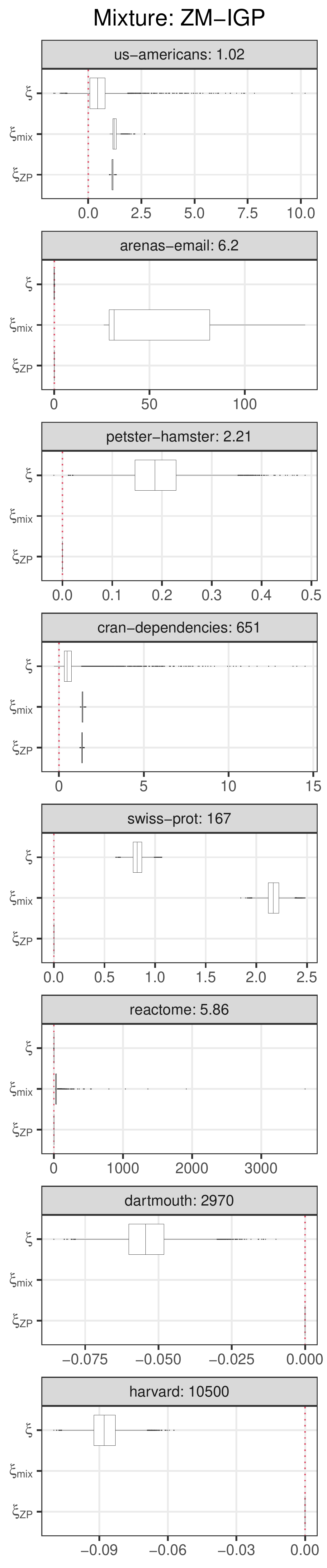} 
\includegraphics[width=0.24\linewidth]{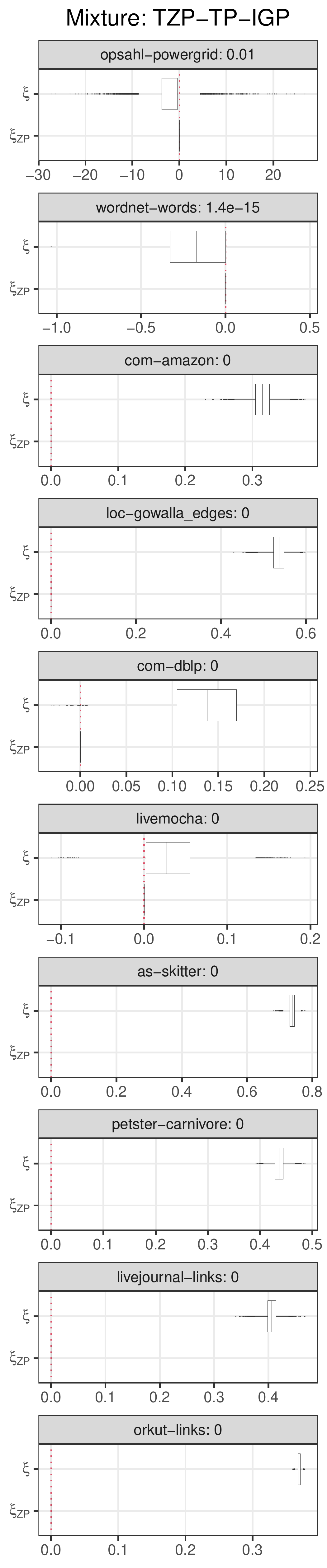} 
\includegraphics[width=0.24\linewidth]{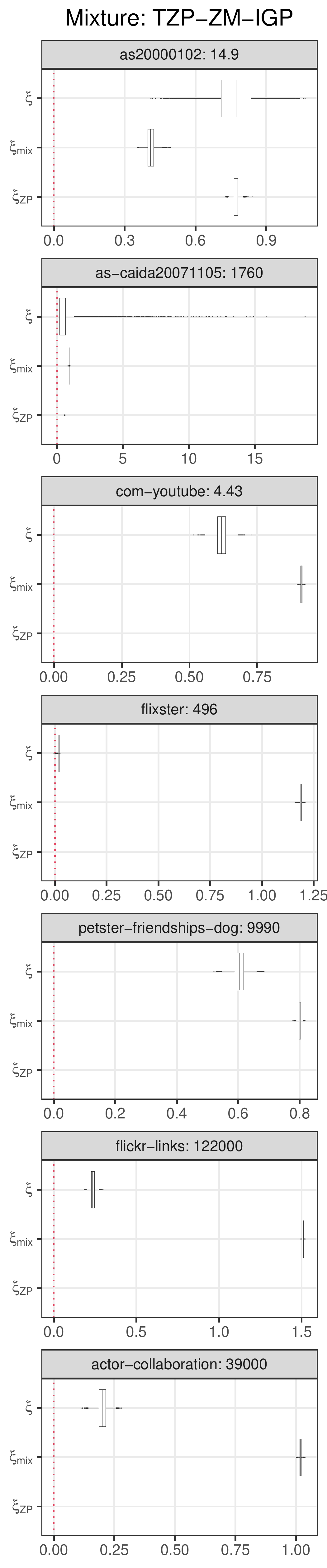} 

}

\caption{Boxplot of $\xi$ and the tail indices implied by the selected mixture and ZP distributions. The number beside the data set name is the Bayes factor (to 3 s.f.) for $\theta_{\text{mix}}=1$ relative to $\theta_{\text{mix}}\in(0,1)$. In the first and third columns, $\xi_{\text{mix}}=0$; in the other two columns, some boxes for $\xi_{\text{mix}}$ are missing as all sampled values of $\alpha_{\text{mix}}$ are smaller than 1.}\label{fig:appendix-plot-xi-alpha}
\end{figure}

\end{knitrout}

\end{document}